\documentclass[sn-mathphys]{sn-jnl}

\usepackage{graphicx}%
\usepackage{multirow}%
\usepackage{amsmath,amssymb,amsfonts}%
\usepackage{amsthm}%
\usepackage{mathrsfs}%
\usepackage[title]{appendix}%
\usepackage{xcolor}%
\usepackage{textcomp}%
\usepackage{manyfoot}%
\usepackage{booktabs}%
\usepackage{algorithm}%
\usepackage{algorithmicx}%
\usepackage{algpseudocode}%
\usepackage{listings}%
\usepackage{bm}
\usepackage{subfig}
\usepackage{caption}
\usepackage{dsfont}
\usepackage{xr}

\DeclareFontFamily{OT1}{pzc}{}
\DeclareFontShape{OT1}{pzc}{m}{it}{<-> s * [1.10] pzcmi7t}{}
\DeclareMathAlphabet{\mathpzc}{OT1}{pzc}{m}{it}

\makeatletter
\newcommand*{\addFileDependency}[1]{
  \typeout{(#1)}
  \@addtofilelist{#1}
  \IfFileExists{#1}{}{\typeout{No file #1.}}
}
\makeatother

\newcommand{\R}{\mathbb{R}}
\newcommand{\vagueconv}{\stackrel{\mathrm{v}}{\rightarrow}} 
\newcommand{\dee}{\mathrm{d}}
\usepackage{stmaryrd}
\usepackage{bm}
\usepackage{bbm}

\DeclareMathOperator*{\argmin}{arg\,min}
\newtheorem{assumption}{Assumption}

\usepackage{geometry}
\usepackage{setspace}
\theoremstyle{thmstyleone}%
\theoremstyle{thmstyletwo}%

\theoremstyle{thmstylethree}%
\newtheorem{definition}{Definition}%

\title[Article Title]{Extreme Value Statistics for Analysing Simulated Environmental Extremes}
\author*[1]{\fnm{Henry} \sur{Elsom}}\email{he294@bath.ac.uk}

\author[1]{\fnm{Matthew} \sur{Pawley}}\email{mtp34@bath.ac.uk}

\affil[1]{\orgdiv{Department of Mathematical Sciences}, \orgname{University of Bath}, \orgaddress{\street{Claverton Down}, \city{Bath}, \postcode{BA2 7AY}, \state{Somerset}, \country{UK}}}

\begin{document}

\begingroup
\singlespacing
\maketitle
\begin{center}
  \large\date{December 20, 2023}  
\end{center}

\vspace{10mm}

\begin{center}
    \textbf{Abstract}\\
    \vspace{5mm}
    \begin{minipage}{0.9\textwidth}
        We present the methods employed by team `Uniofbathtopia' as part of the Data Challenge organised for the 13th International Conference on Extreme Value Analysis (EVA2023), including our winning entry for the third sub-challenge. Our approaches unite ideas from extreme value theory, which provides a statistical framework for the estimation of probabilities/return levels associated with rare events, with techniques from unsupervised statistical learning, such as clustering and support identification. The methods are demonstrated on the data provided for the Data Challenge -- environmental data sampled from the fantasy country of `Utopia' -- but the underlying assumptions and frameworks should apply in more general settings and applications.
    \end{minipage}
\end{center}
\vspace{5mm}
\vfill
\endgroup

\newpage

\section{Introduction}\label{sec1}

The field of environmental sciences has experienced significant advancements, particularly through the utilisation of sophisticated modelling techniques to better understand extreme events. Extreme value analysis is the branch of statistical modelling that focuses on quantifying the frequency and severity of very rare events. Notably, the Peaks over Threshold (PoT) approach \citep{DavSmith1990,pickands75}, which utilises the Generalised Pareto Distribution (GPD), has played a pivotal role in enhancing our comprehension of extreme environmental phenomena. Within climate science, significant strides have been made in the modelling of a broad spectrum of variables, including temperature \citep{clarkson23}, precipitation \citep{katz99}, wind speeds \citep{Kunz2010,FawWalsh06}, as well as other broader environmental topics including hydrology \citep{Towler10,KATZ2002} and air pollution \citep{GOULD2022}.

In this paper, we outline the techniques exercised by the team `Uniofbathtopia' for the Data Challenge organised for the 13$^{\text{th}}$ International Conference on Extreme Value Analysis (EVA2023). A full description of the tasks can be found in the editorial \citep{Rohr23}. We outline our methodologies for each of the four sub-challenges, in which we complement traditional methods from extreme value statistics with other statistical modelling techniques according to the requirements of each task. The challenges involve the estimation of extreme marginal quantiles and marginal/joint exceedance probabilities within the context of an environmental application, designed on the fictitious country of `Utopia'. The organisers of the competition simulated the data using known parameters, so that teams' models could be validated and compared, and in such a way as to to mimic the rich, complex behaviour exhibited by real-world processes. Therefore, we expect that the performance of our proposed methods should extend to general settings and applications.

In the univariate tasks we use approaches from model-based clustering methods including hierarchical models \citep{hastibfri2009} and mixture models \citep{fraley2002}, as well as using Markov Chain Monte Carlo (MCMC) for parameter estimation \citep{coles1996}. We also use bootstrapping methods for confidence interval estimation \citep{gill2020}. For the multivariate problems our approaches are heavily based on the parametric family of max-linear combinations of regularly varying random variables \citep{fougères2013}. We introduce novel methods for performing inference for these models, advancing existing approaches \citep{cooley2019, kiriliouk2022} using modern statistical learning techniques including sparsity-inducing projections and clustering. The two novel aspects of our work include: exploring MCMC parameter estimation bias for small values of the GPD shape parameter, and proposing a new estimator for the noise coefficient matrix of a max-linear model based on sparse projections onto the simplex.

The format of the paper is as follows: Section \ref{univariate} describes the solutions for the univariate challenges with each challenge split into strategy and methodology sections. Section \ref{sec:multivariate} introduces the requisite background theory from multivariate extremes before outlining the methodological frameworks and the results attained for Challenges 3 and 4. We conclude with some final discussion on our solutions in Section \ref{conc}.

\section{Univariate Challenges}\label{univariate}

The first two problems both involve estimating univariate extreme marginal quantiles so we initially describe some of the theory that will be used across both tasks. Suppose that a generalised Pareto distribution (GPD) with scale and shape parameters, \(\sigma\) and \(\xi\) respectively, is a suitable model for exceedances of a threshold \(u\) by a variable \(Y\). Then, for \(y > u\),
\begin{equation}
    \mathbb{P}(Y > y \mid Y > u) = \left(1 + \xi\left(\frac{y-u}{\sigma}\right)\right)_{+}^{-1/\xi}. \nonumber
\end{equation}
Given the probability \(\zeta_u = \mathbb{P}(Y > u)\), the probability of \(Y > y\) can be expressed as,
\begin{equation}
    \mathbb{P}(Y > y) = \zeta_u \left(1 + \xi\left(\frac{y-u}{\sigma}\right)\right)_{+}^{-1/\xi}. \nonumber
\end{equation}
We can find the \(p\)th percentile by rearranging the distribution function to form the quantile function,
\begin{equation}\label{eq:quant}
    q(p \hspace{1mm}; \sigma, \xi, u) = u + \frac{\sigma}{\xi}\left(\left(\frac{p}{\zeta_u}\right)^{-\xi}-1\right).
\end{equation}
However, we can also write down this expression for a return level, that is, the level \(y_T\) that is exceeded on average once every \(T\) years,
\begin{equation}\label{eq:return}
    y_T = u + \frac{\sigma}{\xi}((T n_{\text{yr}} \zeta_u)^{\xi}-1),
\end{equation}
where \(n_{\text{yr}}\) is the number of observations per year. For the purposes of the second challenge \(n_{\text{yr}} = 300\) as we given that a year in Utopia consists of 12 months and 25 days per month.

These expressions imply that we can estimate quantiles and return levels once we obtain estimates of \(\sigma, \xi\) and \(u\). The method of obtaining these values is different for different tasks and will be described, for the univariate challenges, in Sections \ref{sec:chall1} and \ref{sec:chall2}. We also require an estimate of \(\zeta_u\), the probability of an individual observation exceeding the threshold \(u\). We can achieve this by using the empirical probability of \(Y\) exceeding \(u\),
\begin{equation}\label{eq:exceedance}
    \hat{\zeta}_u = \mathbbm{1}\{Y > u\}/n,\nonumber
\end{equation}
that is, the proportion of the total number of observations that exceed the threshold.

\subsection{Challenge 1}\label{sec:chall1}
\subsubsection{Strategy}
For the first task we were asked to calculate extreme quantiles, both point estimates as well as central 50\% confidence intervals. We are interested in the extreme values of the environmental variable \(Y\) which, for the first two univariate tasks, we denote by \(Y_{i}\) for day \(i\). For each day, we also have a vector of covariates \(\bm{X}_i = (V_{1,i}, \dots, V_{8,i})\) with variables \((V_{1}, \dots, V_{4})\) representing unnamed covariates and \((V_{5}, V_{6}, V_{7}, V_{8})\) representing \texttt{Season}, \texttt{Wind Speed}, \texttt{Wind Direction} and \texttt{Atmosphere} respectively. The challenge data is divided into a training set comprising 70 years' worth of daily environmental data and a test set featuring 100 days of environmental data, showcasing diverse combinations of the covariates. 

The first problem requires a model to represent the distribution of \(Y \mid \bm{X}\) in order to estimate the conditional quantiles, \(\{q(0.9999\hspace{1mm};\sigma(\bm{x}_i), \xi(\bm{x}_i), u(\bm{x}_i)) : i = 1, \dots, 100\}\), where
\begin{equation}\label{eq:prob_exp}
    \mathbb{P}(Y \leq q(0.9999\hspace{1mm};\sigma(\bm{x}_i), \xi(\bm{x}_i), u(\bm{x}_i)) \mid \bm{X} = \bm{x}_i) = 0.9999.
\end{equation}
As well as point estimates for the quantiles, it is also necessary to provide central 50\% confidence intervals for these estimates.

We approach this problem by clustering the covariates and analysing the extremes of each cluster individually. We can write the above expression,
\begin{equation}
    \mathbb{P}(Y > q(0.0001\hspace{1mm};\sigma(\bm{x}_i), \xi(\bm{x}_i), u(\bm{x}_i)) \mid Y > u,\bm{X} = \bm{x}_i) \mathbb{P}(Y>u \mid \bm{X} = \bm{x}_i)  = 0.0001, \nonumber
\end{equation}
where a GPD can be used to describe the first probability in the expression. 

Our approach was to derive clusters for the covariate values. We make the assumption that days which have similar values of the environmental covariates will generate extreme values with very similar marginal tail behaviour that can be modelled by a GPD with the same scale and shape parameters. To represent the cluster structure, we introduce \(N\) latent random variables \(\bm{Z} = (Z_1, \dots, Z_N)\) where \(N\) is the total number of daily observations in the dataset. Let \(J \in \{1, \dots, K\}\) denote the number of clusters. Then, for each \(i = 1, \dots, N\), \(Z_i = j\) corresponds to the \(i\)-th observation being allocated to the \(j\)-th cluster. The probability of belonging to cluster \(j\) given the covariates \(\bm{x}_i\) can be estimated by first fitting a mixture of multivariate normal distributions to the covariates in the training dataset. The mixture model is subsequently applied to determine the probabilities \( \mathbb{P}(Z_i = 1 \mid \bm{X} = \bm{x}_i), \ldots, \mathbb{P}(Z_i = J \mid \bm{X} = \bm{x}_i)\) with,
\begin{equation}
    \mathbb{P}(Z_i = j \mid \bm{X} = \bm{x}_i) = \frac{\mathbb{P}(Z_i = j)  \mathbb{P}(\bm{X} = \bm{x}_i \mid Z_i = j)}{\sum^K_{j=1} \mathbb{P}(Z_i = j) \mathbb{P}(\bm{X} = \bm{x}_i \mid Z_i = j)}, \nonumber
\end{equation}
where \(\mathbb{P}(Z_i = j)\) is the mixing component and \(\mathbb{P}(\bm{X} = \bm{x}_i \mid Z_i = j)\) is the multivariate normal distribution. The probabilities are defined by the estimated parameters for the mixture of multivariate normal distributions. We now rewrite (\ref{eq:prob_exp}) to form our model for the probability of \(Y > q(0.0001\hspace{1mm};\sigma(\bm{x}_i), \xi(\bm{x}_i))\) conditional on \(Z_i = j\) and \(\bm{X} = \bm{x}_i\),
\begin{align}\label{eq:probs}
    \mathbb{P}(Y > q(0.0001\hspace{1mm}; &\sigma(\bm{x}_i), \xi(\bm{x}_i)) \mid Z_j, \bm{X}) \nonumber \\ &= \mathbb{P}(Y > q(0.0001\hspace{1mm};\sigma(\bm{x}_i), \xi(\bm{x}_i)) \mid Z, Y > u_j) \mathbb{P}(Y > u_j | Z), \nonumber
\end{align}
where \(u_j\) is the threshold set for all observations within cluster \(j\).
In order to provide central 50\% confidence intervals for these estimates, we use a jackknife (leave-one-out) resampling method which will be described in greater detail in the following section.

\subsubsection{Methodology}\label{sec:q1meth}
We start by clustering the covariates using the method by \cite{fraley2003}. As such, we assume that the covariates are sampled from a mixture of multivariate normal distributions and we use \(\phi(\cdot \mid \bm{\mu}, \bm{\Sigma})\) to denote the probability density function of the Gaussian distribution with mean vector \(\bm{\mu}\) and covariance matrix \(\bm{\Sigma}\). More formally, a \(d\)-dimensional Gaussian mixture model (GMM) with \(K\) components can be specified by a collection \(\bm{\mu} = \{\bm{\mu}_1, \dots, \bm{\mu}_K\}\) of \(d\)-dimensional mean vectors, a vector \(\bm{\alpha} = \{\alpha_1, \dots, \alpha_K\}\) of mixture probabilities, where we enforce \(\alpha_j > 0\) and \(\sum^K_{j=1}\alpha_j = 1\), and the covariance matrices are given by  \(\bm{\Sigma} = \{\bm{\Sigma}_1, \dots, \bm{\Sigma}_K\}\). The conditional probability density function can be written,
\begin{equation}\label{mixture}
    \mathbb{P}(\bm{X} = \bm{x}_i \mid \mathbf{\Psi}) = \sum^K_{j=1}\alpha_j \phi(\bm{x};\bm{\mu}_j, \bm{\Sigma}_j), \nonumber
\end{equation}
where \(\mathbf{\Psi} = \{\bm{\alpha}, \bm{\mu}, \bm{\Sigma}\}\) are the parameters of the mixture model.

\begin{figure}
  \subfloat[Season 1]{\includegraphics[width=0.5\textwidth]{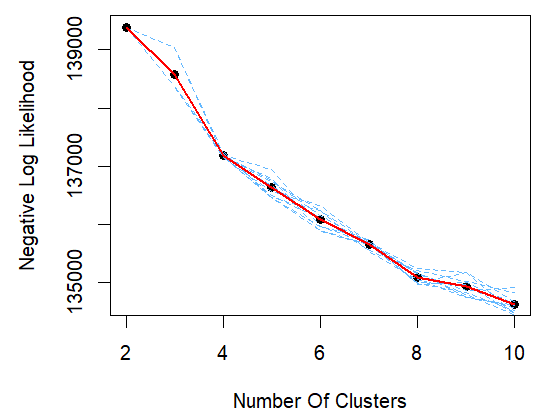}\label{fig:sea1}}
  \hfill
  \subfloat[Season 2]
  {\includegraphics[width=0.49\textwidth]{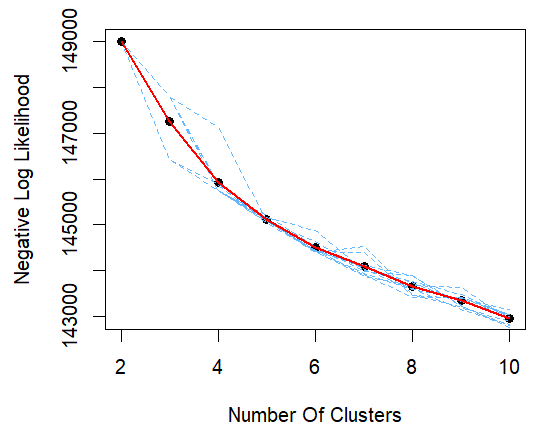}\label{fig:sea2}}
  \caption{An elbow plot to determine an optimum number of clusters for each season. The figure displays the negative log-likelihood for different simulations (blue) as well as the average (red).}
  \label{fig:elbow}
\end{figure}

We performed clustering separately for each of the two seasons - Season 1 and Season 2. The optimum number of clusters was then found by using the elbow method; we derive the negative log-likelihood for each choice of \(J\), and identify the point that this value begins to plateau. These results are displayed in \autoref{fig:elbow}. For both seasons the optimum number of covariate clusters was determined to be \(J = 5\) for each season. There was some flexibility on this value, however, we recognised the trade off between maximising the number of covariate clusters and maximising the number of data points within each cluster.
Assuming that we have decided on an appropriate fixed \(J\) for each season, the mixture model parameters \(\bm{\Psi}\) are unknown and to be found. In order to achieve this, we use the Expectation Maximisation (EM) algorithm to perform maximum likelihood estimation \citep{mclachlan2000}, which is executed using the \texttt{mclust} package in \texttt{R}. Once the covariates were split into their respective clusters, it was possible to collect the values of \(Y\) within each corresponding cluster, denoted \(\mathcal{Y}^{(j)} = \{Y_i : Z_i = j\}\).

We were then able to perform extreme value analysis on each set of points \(\mathcal{Y}^{(j)}\) separately, by fitting a GPD to the extremal data, as outlined at the start of Section \ref{univariate}. Each of the thresholds, \(u_j\), were found by interpreting mean residual life plots and parameter stability plots \citep{Coles2001}. Once the GPD parameters were found for each cluster, it was possible to then substitute these values into the quantile function (\ref{eq:quant}). This is used to calculate the quantile point estimate for each cluster. 
The challenge also requires central 50\% confidence intervals for the estimates of these extreme conditional quantiles. For each set of points in a cluster, \(\mathcal{Y}^{(j)}\) we are able to find a estimate of the quantile by performing a jackknife resampling of the extremal data and using this to fit the GPD. 

In the jackknife resampling method, you systematically create a set of samples by leaving out one observation at a time and calculating the quantile point estimate based on the remaining observations. For each observation \(i\) in the extremal dataset \(\mathcal{Y}^{(j)} \mid \mathcal{Y}^{(j)} > u_j\) create a new dataset \(\mathcal{Y}^{(j)}_{(-i)} \mid \mathcal{Y}^{(j)} > u_j\) by excluding the \(i\)-th observation. You then fit a new GPD to this data and, using these parameter estimates, calculate the quantile point estimate in (\ref{eq:quant}). Once you have undertaken this process for every observation in the dataset you calculate the empirical 25th and 75th percentiles for the jackknife samples.

The performance of this method was disappointing, we think for two reasons. There were no clearly defined clusters. This points to the fact that the fitted Gaussian mixture distribution may be unable to fully capture the non-Gaussian distribution of the covariate values. In \autoref{fig:elbow}, the point at which the negative log-likelihood levels off is really not clear, illustrating this argument. The second reason maybe down to the fact that covariates were not used when fitting the GPDs to the extreme data in each cluster, only for cluster assignment. A better GPD model for each cluster may have resulted in better quantile estimates.

\subsection{Challenge 2}\label{sec:chall2}
\subsubsection{Strategy}
\begin{figure}
    \centering
    \includegraphics[width=\linewidth]{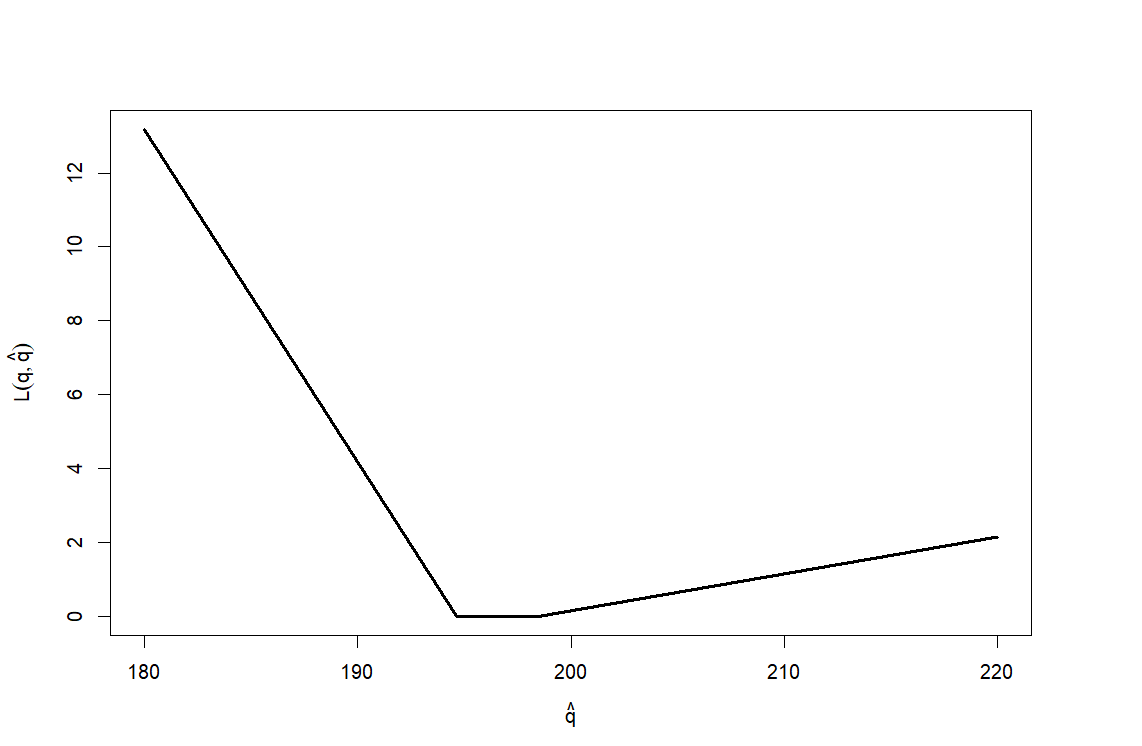}
    \caption{A diagram of the loss function using the quantile value \(q = 196.6\).}
    \label{fig:losfunc}
\end{figure}

The second challenge is to estimate the marginal quantile \(q\) such that
\begin{equation}
    \mathbb{P}(Y > q) = \frac{1}{300 T}, \nonumber
\end{equation}
where \(T=200\), that is, an estimate of the return level of the 1 in 200 year event for Amaurot. This challenge is assessed using a loss function that penalises under-estimating the quantile more than over-estimating it. Formally, for a given estimate \(\hat{q}\) and the true marginal quantile \(q\), the loss is calculated as,
\begin{equation}\label{eq:lossy}
    L(q,\hat{q}) = \begin{cases}
        0.9(0.9q-\hat{q}) & \text{if } 0.99q > \hat{q},\\
        0,  & \text{if } |q - \hat{q}| \leq 0.01q,\\
        0.1(\hat{q} - 1.01q), & \text{if }1.01q<\hat{q}.
    \end{cases}
\end{equation}
This is to replicate a real world case, for example, over-estimating flood defences means increased cost, under-estimating flood defences has more draconian consequences like fatalities. After completing the challenge, we were given the correct quantile \(q = 196.6\) and so we can substitute this value into the loss function, as displayed in \autoref{fig:losfunc}.

As described in Section \ref{univariate}, we can estimate the return level by assuming that a suitable model for the extremal data is a GPD with scale and shape parameters, \(\sigma\) and \(\xi\) for exceedances of threshold \(u\) from (\ref{eq:return}),
\begin{equation}\label{eq:ret}
    y_{200} = u + \frac{\sigma}{\xi}((200 \times 300 \zeta_u)^{\xi}-1),
\end{equation}
where we have substituted \(T = 200\) and \(n_y = 300\) as we are calculating the 200 year return level where we have 300 daily observations a year.
\begin{figure}[t]\centering
\subfloat[Survival Curve]{\label{a}\includegraphics[width=.95\linewidth]{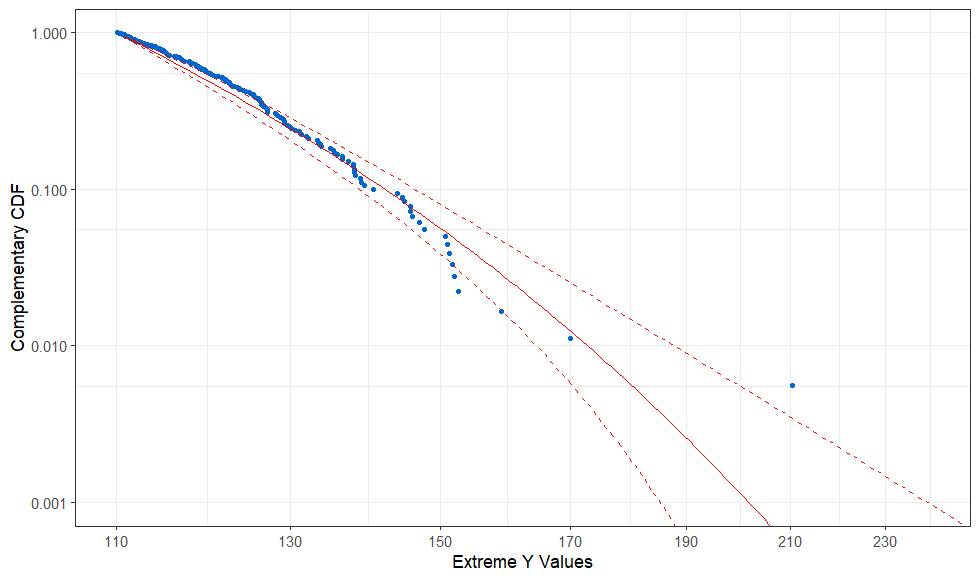}}\par
\subfloat[Scale Parameter]{\label{b}\includegraphics[width=.48\linewidth]{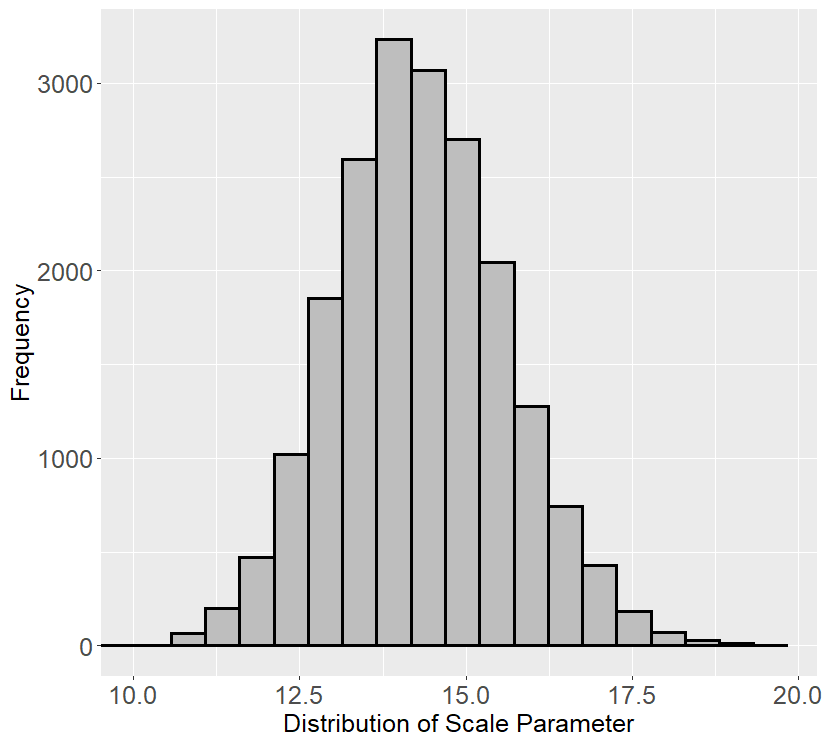}}
\subfloat[Shape Parameter]{\label{c}\includegraphics[width=.48\linewidth]{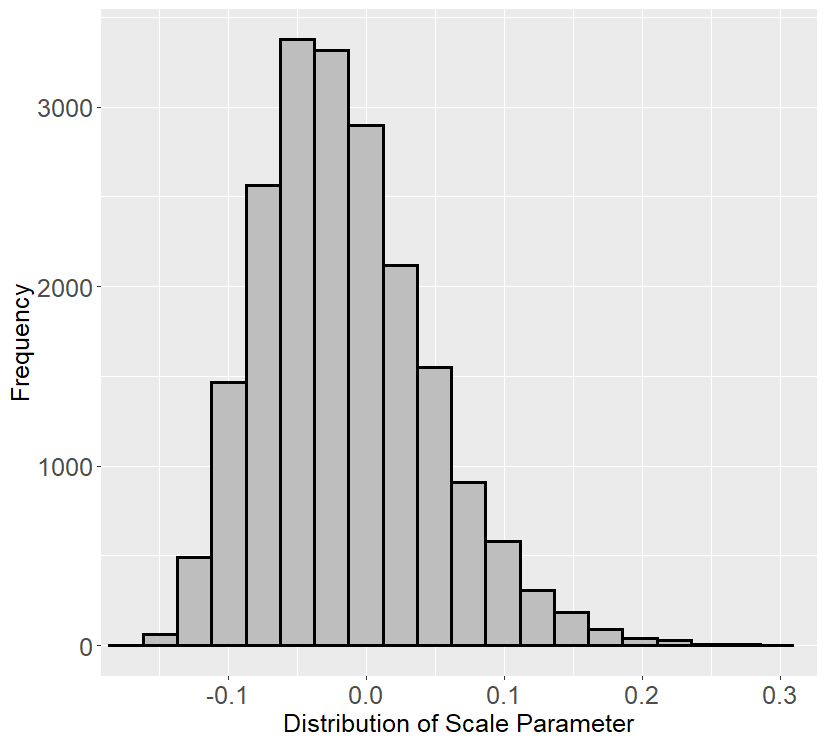}}\hfill
\caption{A Bayesian fit to the survival curve for extreme \(Y\) values above threshold 110, with distributions of sampled scale and shape parameters.}
\label{fig:survive}
\end{figure}
We used Markov Chain Monte Carlo (MCMC) methods to find an initial GPD estimate for the extremal data, as displayed by the survival curve in \autoref{fig:survive}. There were 180 observations above the chosen threshold \(u=110\). After observing the values of the posterior samples of the scale and shape parameters, we explored whether there was a bias in the parameter estimates that arises when applying our MCMC framework to data generated from a GPD with a small (positive or negative) shape parameter and a large scale parameter. We create 1000 sets of simulated data of 180 points from GPDs with different random scale and shape parameters. The scale and shape parameters are drawn from uniform distributions, and the boundaries of these distributions are determined by examining the histograms of the parameters sampled during the initial survival curve fitting process. We then use the same Bayesian fitting procedure as before and find the difference between the simulated parameters and the estimated parameter values.

\subsubsection{Methodology}
We estimate the parameters using a Bayesian model, generating estimations of \(y_{200}\) from samples from the posterior distribution. We assume \(m\) extreme excesses over a threshold \(u\), follow the model,
\begin{equation}
    Y-u \mid Y > u \sim \text{GPD}(\sigma, \xi), \nonumber
\end{equation}
where \(u\) is determined using a mean excess plot. The shape and scale parameters are assumed to be independent a priori. The likelihood function is the product of the density function values for each observation,
\begin{equation}
    \mathcal{L}(\sigma, \xi | y) = \prod^m_{i=1} \frac{1}{\sigma}\left(1+\frac{\xi y_i}{\sigma}\right)^{-(1/\xi+1)}, \nonumber
\end{equation}
\begin{figure}
  \subfloat[Scale Parameter]{\includegraphics[width=0.5\textwidth]{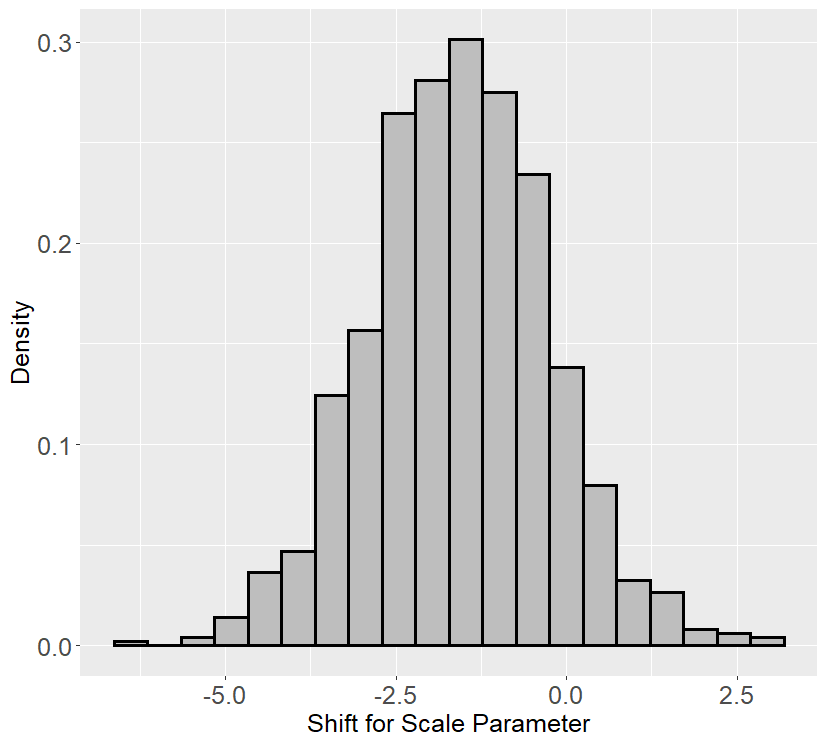}\label{fig:scalecorrect}}
  \hfill
  \subfloat[Shape Parameter]{\includegraphics[width=0.5\textwidth]{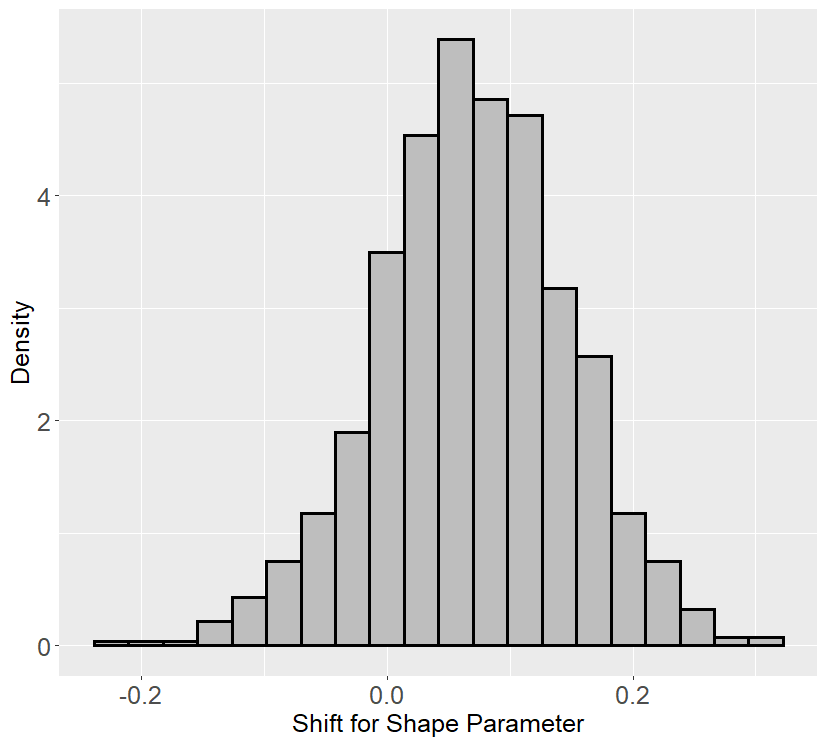}\label{fig:shapecorrect}}
  \caption{Histogram of residuals of predicted and simulated scale (left) and shape (right) parameters.}
  \label{fig:corrs}
\end{figure}
and we choose prior distributions for \(\sigma\) and \(\xi\),
\begin{align}
    \sigma \sim & \text{ Gamma}(4, 1) \nonumber \\
    \xi \sim & \hspace{0.5mm}\text{ Normal}(0, 1). \nonumber
\end{align}

We fit a survival curve to \(y\) with threshold \(u = 110\) and show the resulting shape and scale distributions. These histograms are displayed in \autoref{fig:survive}. Once we have these distributions it is then possible to plot the return levels using (\ref{eq:ret}). There were 180 points above the threshold out of a possible 21000 data points and so we use the empirical estimate for the exceedance probability \(\hat{\zeta}_u = 180/ 21000 \approx 0.0086\).

We then create 1000 sets of simulated data of 180 points from GPDs with different random scale and shape parameters. Formally, we propose the total set of simulated data \(\tilde{\mathpzc{y}} = \{\tilde{y}^{(1)}, \tilde{y}^{(2)}, \dots, \tilde{y}^{(1000)}\}\), and the set of random scale and shape parameters \(\tilde{\bm{\Sigma}} = \{\tilde{\sigma}^{(1)}, \tilde{\sigma}^{(2)}, \tilde{\sigma}^{(1000)}\}\), \(\tilde{\bm{\Xi}} = \{\tilde{\xi}^{(1)}, \tilde{\xi}^{(2)}, \tilde{\xi}^{(1000)}\}\) respectively. Each \(\tilde{y}^{(i)}\) is 180 data points, \(\tilde{y}^{(i)} = \{\tilde{y}^{(i)}_1, \tilde{y}^{(i)}_2, \dots, \tilde{y}^{(i)}_{180}\}\)  sampled from \(\tilde{y}^{(i)} \sim \text{GPD}(\tilde{\sigma}^{(i)}, \tilde{\xi}^{(i)})\) and where \(\tilde{\sigma}^{(i)} \sim \text{Unif}(11, 18)\) and \(\tilde{\xi}^{(i)} \sim \text{Unif}(-0.15, 0.20)\). These limits are taken from the distribution limits of the samples from the survival curve fit.
For each dataset we assume the model,
\begin{equation}
    \tilde{y}^{(i)} \sim \text{GPD}(\tilde{\sigma}^{(i)}, \tilde{\xi}^{(i)}), \nonumber
\end{equation}
\begin{figure}[t]
     \centering
     \includegraphics[width=\linewidth]{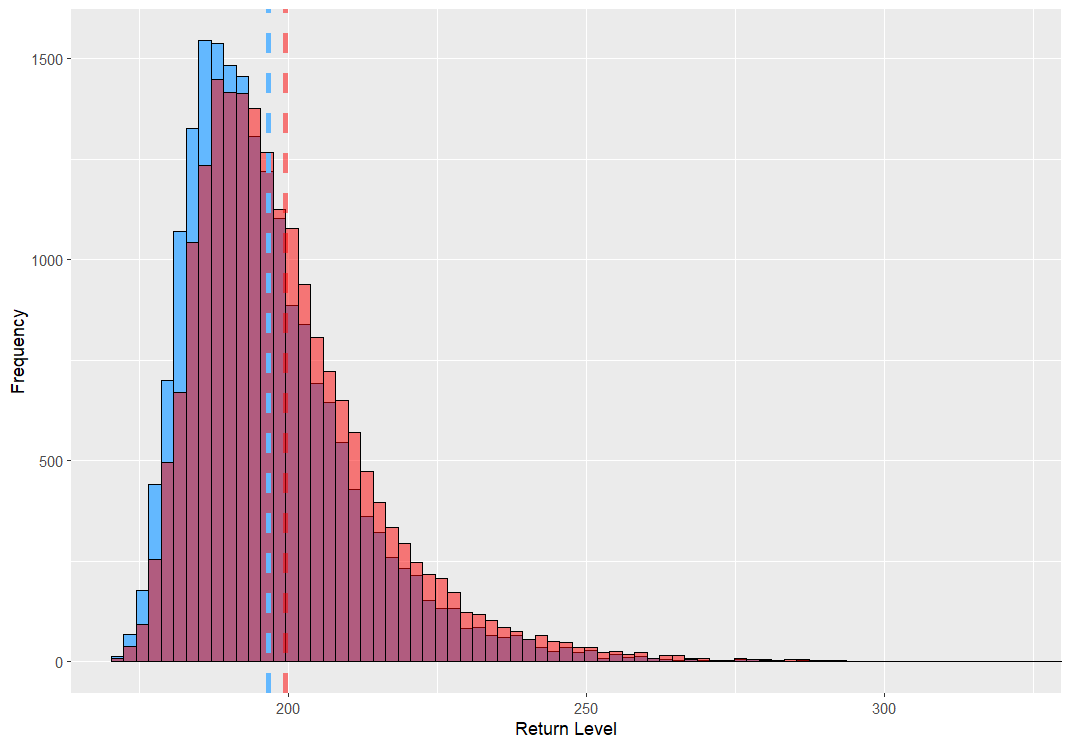}
     \caption{The return level samples before shifting (blue) and after shifting (red). The dashed lines denote the mean of the samples.}
     \label{fig:returns}
 \end{figure}and fit the survival function as before. We keep the threshold as \(u = 110\). Each time the model is fit to a new simulated dataset, the estimated parameters are recorded. It is then possible to find the difference between the estimated parameters and the simulated parameters to estimate the bias. These distributions are displayed in \autoref{fig:corrs}. We then apply the mean corrections to the initial return level samples and compare the mean return levels. These are shown in \autoref{fig:returns}. This gives an overall increase in the average return level from 196.4 to 199.4.
The correct quantile value was \(q = 196.6\). Although this bias correction actually increased the loss from \(L(196.6, 196.4) = 0\) to \(L(196.6, 199.4) = 0.0834\) (using (\ref{eq:lossy})), it was reassuring that the bias correction shifted our initial underestimate to an overestimate.

\section{Multivariate Challenges}\label{sec:multivariate}

In Challenges 3 and 4 we are presented with a $d$-dimensional random vector $\bm{X}=(X_1,\ldots,X_d)\sim F_{\bm{X}}$, representing the value of an environmental variable at $d$ sites in Utopia, where $F_{\bm{X}}$ is an unknown joint distribution function. Our goal is to estimate the probability that $\bm{X}$ lies in a given extreme `failure region' based on a sample of independent observations of $\bm{X}$. The failure regions of interest take the form $\{\bm{X}_\beta > \bm{u}, \bm{X}_{\beta^c} < \bm{l}\}$, where $\beta\subset\{1,\ldots,d\}$, $\bm{u}$ is a vector of high thresholds, and $\bm{l}$ is vector of upper limits of lower order.
The inherent difficulty of the task stems from the events' return periods being of similar order or even significantly longer than the observation period over which the data are collected; empirical methods based solely on the relative frequency of event occurrences are fruitless. Instead, we use the observed data to infer an estimate for the probabilistic structure of the tail of $\bm{X}$ and subsequently compute the tail event probability under this model. This encapsulates the philosophy of multivariate extreme value statistics. 

In the absence of prior knowledge about the physical processes driving the environment of Utopia, we are compelled to recourse to data-driven, statistical learning methods for multivariate extremes. The particular tools we choose will depend on the particular nature of each challenge and the difficulties it presents. In Challenge 3, the failure regions are defined by \textit{different} subsets of variables being large, thereby placing emphasis on accurately modelling the so-called extremal directions of $\bm{X}$. Challenge 4 is a high-dimensional problem, which calls for the use of dimension reduction techniques, such as clustering, in order to overcome the curse of dimensionality inherent to tail dependence estimation.

All the code for our Challenge 3 and 4 methodologies is available at \url{https://github.com/pawleymatthew/eva-2023-data-challenge}.

\subsection{Background}
\subsubsection{Multivariate regular variation and the angular measure}

As is common in multivariate extremes, our approach will be based on the framework of multivariate regular variation (MRV). We present two formulations of the MRV property that will be useful for us. We shall assume that $\bm{X}$ takes values on the positive orthant $\R_+^d:=[0,\infty)^d$.

\begin{definition}\label{def:mrv}
    We say that $\bm{X}$ is \textit{multivariate regularly varying} with index $\alpha>0$, denoted $\bm{X}\in\mathrm{RV}_+^d(\alpha)$, if it satisfies the following equivalent definitions \citep[Theorem 6.1]{resnick2007}:
    \begin{enumerate}
        \item There exists a sequence $b_n\to\infty$ and a non-negative Radon measure $\nu_{\bm{X}}$ on $\mathbb{E}_0:=[0,\infty]^d\setminus\{\bm{0}\}$ such that
        \begin{equation}\label{eq:mrv}
            n\mathbb{P}(b_n^{-1}\bm{X} \in \cdot) \vagueconv \nu_{\bm{X}}(\cdot),\qquad (n\to\infty),
        \end{equation}
        where $\vagueconv$ denotes vague convergence in the space of non-negative Radon measures on $\mathbb{E}_0$. The \textit{exponent measure} $\nu_{\bm{X}}$ is homogeneous of order $-\alpha$, i.e. $\nu_{\bm{X}}(s\,\cdot)=s^{-\alpha}\nu_{\bm{X}}(\cdot)$ for any $s>0$.
        \item For any norm $\|\cdot\|$ on $\R^d$, there exists a sequence $b_n\to\infty$ and a finite measure $H_{\bm{X}}$ on $\mathbb{S}_+^{d-1}:=\{\bm{x}\in\R_+^d:\|\bm{x}\|=1\}$ such that for $(R,\bm{\Theta}):=(\|\bm{X}\|,\bm{X}/\|\bm{X}\|)$,
        \begin{equation}\label{eq:polar-mrv}
            n\mathbb{P}((b_n^{-1}R,\bm{\Theta}) \in \cdot) \vagueconv \nu_{\alpha}\times H_{\bm{X}}(\cdot),\qquad (n\to\infty),
        \end{equation}
        in the space of non-negative Radon measures on $(0,\infty]\times\mathbb{S}_+^{d-1}$, where $\nu_{\alpha}((x,\infty))=x^{-\alpha}$ for any $x>0$. The \textit{angular measure} $H_{\bm{X}}$ has total mass $m:=H_{\bm{X}}(\mathbb{S}_+^{d-1})\in(0,\infty)$.
    \end{enumerate}
\end{definition}

The limit measures $\nu_{\bm{X}}$ and $H_{\bm{X}}$ are related via 
\begin{align*}
    \nu_{\bm{X}}(\{\bm{x}\in\mathbb{E}_0:\|\bm{x}\|>s,\bm{x}/\|\bm{x}\|\in\cdot\}) 
    &= s^{-\alpha}H_{\bm{X}}(\cdot),\\ 
    \nu_{\bm{X}}(\dee r\times \dee \bm{\theta})
    &=\alpha r^{\alpha-1}\dee r\,\dee H_{\bm{X}}(\bm{\theta}).
\end{align*}
The MRV property means that the probabilistic tail of $\bm{X}$ can be decomposed into a univariate $\alpha$-regularly varying radial component \citep[Theorem 3.6]{resnick2007} that is asymptotically independent of the angular component. The angular measure represents the limiting distribution of the angular component and encodes all information about the tail dependence structure. 

Henceforth, whenever we encounter a random vector $\bm{X}\in\mathrm{RV}_+^d(\alpha)$, we shall assume that its marginal components are identically distributed with a Fréchet distribution with shape parameter $\alpha$, that is $\mathbb{P}(X_i<x)=\Phi_{\alpha}(x):=\exp(-x^{-\alpha})$ for $i=1,\ldots,d$. Moreover, we will always choose $\|\cdot\|=\|\cdot\|_\alpha$, the $L_\alpha$-norm on $\R^d$, and specify that the normalising sequence in \eqref{eq:polar-mrv} is $b_n=n^{1/\alpha}$. With these particular choices the marginal variables have unit scale (see Definition 4 in \cite{klüppelberg2021}) and $H_{\bm{X}}$ has mass $m=d$.

The problem of modelling the angular measure has received considerable attention in recent years. One avenue of research concerns learning the support of the angular measure, or equivalently, the sets of variables that may be concurrently extreme. Let $\mathcal{P}_d^\star$ denote the power set of the index set $\mathbb{V}_d:=\{1,\ldots,d\}$ excluding the empty set. A set $\beta\in\mathcal{P}_d^\star$ is termed an extremal direction of $\bm{X}\in\mathrm{RV}_+^d(\alpha)$ if the subspace
    \begin{equation*}
    C_\beta = \{\bm{w}\in\mathbb{S}_+^{d-1} : w_i > 0 \,\, \forall i\in\beta, \, w_i=0 \,\, \forall i\in\beta^c \} \subseteq \mathbb{S}_+^{d-1}
\end{equation*}
has non-zero $H_{\bm{X}}$-mass. Another research direction tackles the challenge of modelling the angular measure in high dimensions. One approach is to consider a summary of the full dependence structure encoded in a matrix of pairwise extremal dependence metrics. One such matrix, popularised by \cite{cooley2019}, is the tail pairwise dependence matrix (TPDM). The TPDM of $\bm{X}\in\mathrm{RV}_+^d(2)$ is the $d\times d$ matrix $\Sigma_{\bm{X}}=(\sigma_{\bm{X}_{ij}})$ with entries
\begin{equation}\label{eq:tpdm}
    \sigma_{\bm{X}_{ij}} = \int_{\mathbb{S}_+^{d-1}} \theta_i\theta_j \,\dee H_{\bm{X}}(\bm{\theta}), \qquad (i,j=1,\ldots,d).
\end{equation}
The diagonal entries are the squared margin scales; in our case $\sigma_{\bm{X}_{ii}}=1$ for $i=1,\ldots,d$. The off-diagonal entries relate to the extremal dependence between the associated pairs of variables \citep{resnick2004}. In particular, $\sigma_{\bm{X}_{ij}}=0$ if and only if $X_i$ and $X_j$ are asymptotically independent. For asymptotically dependent variables the magnitude of $\sigma_{\bm{X}_{ij}}$ measures the strength of extremal dependence. Cruciually for us, the TPDM is completely positive \citep[Proposition 5]{cooley2019}.
\begin{definition}
    A matrix $S$ is \textit{completely positive} if there exists a matrix $A$ with non-negative entries such that $S=AA^T$. We call $AA^T$ a \textit{CP-decomposition} of $S$ and $A$ a \textit{CP-factor} of $S$.
\end{definition}
This property will connect the TPDM to the parametric family of multivariate extreme value models introduced in the following section. 

\subsubsection{The max-linear model for multivariate extremes}\label{sec:max-linear-model}

Our proposed methods for Challenges 3 and 4 make use of the \textit{max-linear model}, a parametric model based on the class of random vectors constructed by max-linear combinations of independent Fréchet random variables \citep{fougères2013}. This model appealed to us for several reasons. First, it is flexible in the sense that any regularly varying random vector can be arbitrarily well-approximated by a max-linear model with sufficiently many parameters \citep{fougères2013}. Since neither Challenge 3 nor Challenge 4 provide any prior information about the underlying data-generating processes, it is desirable to avoid imposing overly restrictive assumptions on the tail dependence structure. Secondly, although the number of parameters grows rapidly -- at least $\mathcal{O}(d)$ but often even $\mathcal{O}(d^2)$ -- efficient inference procedures are available even in high dimensions. Scalability is critical for Challenge 4, where $d=50$. Finally, extremal directions/failure probabilities can be straightforwardly identified/approximated from the model parameter \citep{kiriliouk2022}, facilitating the computations necessary for Challenges 3 and 4.

\begin{definition}
    For some $q\geq 1$, let $\bm{Z}=(Z_1,\ldots,Z_q)$ be a random vector with $Z_1,\ldots,Z_q\sim\Phi_{\alpha}$ independently and $A=(a_{ij})\in\R_{+}^{d\times q}$. The random vector $\bm{X}=(X_1,\ldots,X_d)$ with components
    \begin{equation*}
        X_i := \bigvee_{j=1}^q a_{ij}Z_j,\qquad (i=1,\ldots,d)
    \end{equation*}
is said to be \textit{max-linear} with parameter matrix $A$, denoted $\bm{X}\sim\mathrm{MaxLinear}(A,\alpha)$. We call $A$ the \textit{noise coefficient matrix}, $\bm{Z}$ the \textit{innovations vector}, and write $\bm{X}=A\times_{\max} \bm{Z}$.
\end{definition}
 \cite{cooley2019} show that $\bm{X}=A\times_{\max}\bm{Z}\in\mathrm{RV}_+^d(\alpha)$ and its angular measure is
\begin{equation}\label{eq:max-linear-H}
    H_{\bm{X}}(\cdot) = \sum_{j=1}^q \|\bm{a}_j\|_{\alpha}^{\alpha} \delta_{\bm{a}_j/\|\bm{a}_j\|_\alpha}(\cdot).
\end{equation}
The angles along which extremes can occur are (in the limit) precisely the self-normalised columns of $A$. Substituting \eqref{eq:max-linear-H} into \eqref{eq:tpdm}, we observe that $\bm{X}\sim\mathrm{MaxLinear}(A,\alpha=2)$ has TPDM $\Sigma_{\bm{X}}=AA^T$. In other words, the noise coefficient matrix is a CP-factor of the model TPDM. Conversely, given an arbitrary random vector $\bm{X}\in\mathrm{RV}_+^d(2)$ with TPDM $\Sigma_{\bm{X}}$, any CP-factor $A$ of $\Sigma_{\bm{X}}$ parametrises a max-linear model with identical pairwise dependence metrics to $\bm{X}$, i.e. $\Sigma_{A\times_{\max}\bm{Z}}=\Sigma_{\bm{X}}$.

\cite{kiriliouk2022} give examples of classes of tail events $\mathcal{C}\subset\mathbb{E}_0$ for which $\mathbb{P}(\bm{X}\in\mathcal{C})$ can be approximated by a function of the parameter $A$. Considering Challenges 3 and 4, we focus on regions associated with events where $\bm{X}$ is large only in the set of components indexed by $\beta=\{\beta_1,\ldots,\beta_s\}\in\mathcal{P}_d^\star$, for $1\leq s\leq d$. Such a region is defined by
\begin{equation*}
    \mathcal{C}_{\beta,\bm{u}} := \{\bm{x}\in \mathbb{E}_0:\bm{x}_{\beta} >\bm{u}, \bm{x}_{\beta^c} < \bm{l} \},\qquad \mathbb{P}(\bm{X}\in \mathcal{C}_{\beta,\bm{u}}) = \mathbb{P}(\bm{X}_\beta > \bm{u},\bm{X}_{\beta^c}<\bm{l}),
\end{equation*}
where $\bm{u}=(u_1,\ldots,u_s)\in\R_+^s$ is a vector of high thresholds, and $\bm{l}\in\R_+^{d-s}$ is vector of comparatively small upper thresholds. An approximate formula for $\mathbb{P}(\bm{X}\in \mathcal{C}_{\beta,\bm{u}})$ in the case $\beta=\mathbb{V}_d$ is stated in Section 2.3 of \cite{kiriliouk2022}. We generalise this result to arbitrary $\beta$ as follows: if $\bm{X}\sim\mathrm{MaxLinear}(A,\alpha)$ with $A=(a_{ij})\in\R_+^{d\times q}$ and $\min_{i=1,\ldots,s} u_i$ is sufficiently large, then
\begin{equation}\label{eq:prob-approx-formula}
    \mathbb{P}(\bm{X}\in\mathcal{C}_{\beta,\bm{u}}) \approx \hat{\mathbb{P}}(\bm{X}\in\mathcal{C}_{\beta,\bm{u}}) := \sum_{j:\frac{\bm{a}_j}{\|\bm{a}_j\|_\alpha}\in C_\beta} \min_{i=1,\ldots,s}\left(\frac{a_{\beta_i,j}}{u_i}\right)^\alpha.
\end{equation}
If $\bm{u}=u\bm{1}_s$ for some large scalar $u>0$, then we write $\mathcal{C}_{\beta,u}:=\mathcal{C}_{\beta,u\bm{1}_s}$. We derive the formula for this simpler case but the steps can be easily modified for general $\bm{u}$.
From \eqref{eq:mrv} we have that, provided $u$ is sufficiently large,
\begin{equation*}
    \mathbb{P}(\bm{X}\in\mathcal{C}_{\beta,u})
    = \frac{1}{n}\left[n\mathbb{P}\left(\frac{\bm{X}}{n^{1/\alpha}}\in \frac{\mathcal{C}_{\beta,u}}{n^{1/\alpha}}\right)\right]
    \approx \frac{1}{n}\nu_{\bm{X}}\left(\frac{\mathcal{C}_{\beta,u}}{n^{1/\alpha}}\right)
    =\nu_{\bm{X}}(\mathcal{C}_{\beta,u}),
\end{equation*}
where the last step follows from homogeneity of the exponent measure. Transforming to pseudo-polar coordinates we have
\begin{equation}\label{eq:integral-formula-proof}
    \nu_{\bm{X}}(\mathcal{C}_{\beta,u})
    = \int_{\left\lbrace (r,\bm{w})\,:\,r\bm{w}_\beta>u\bm{1}_{s},\,r\bm{w}_{\beta^c}<l\bm{1}_{d-s}\right\rbrace} \alpha r^{-\alpha-1}\,\dee r\,\dee H_{\bm{X}}(\bm{w}).
\end{equation}
Based on \eqref{eq:max-linear-H}, there are only $q$ possible angles along which extremes can occur, namely $\bm{a}_j/\|\bm{a}_j\|_\alpha$ for $j=1,\ldots,q$. However, only those angles $\bm{a}_j/\|\bm{a}_j\|_\alpha\in C_\beta$ will contribute to the integral. If $\bm{w}=\bm{a}/\|\bm{a}\|_\alpha\not\in C_\beta$, then there exists $i\in\{1,\ldots,s\}$ such that $w_{\beta_i}=0$ and hence $rw_{\beta_i}=0 < u$ for all $0<r<\infty$. On the other hand, if $\bm{w}=\bm{a}/\|\bm{a}\|_\alpha\in C_\beta$, then $\bm{w}_{\beta^c}=\bm{0}\in\R^{d-s}$ and so
\begin{equation*}
    r\bm{w}_\beta>u\bm{1}_{s},\,r\bm{w}_{\beta^c}<l\bm{1}_{d-s} 
    \iff r\bm{w}_\beta>u\bm{1}_{s}
    \iff r > \max_{i=1,\ldots,s}\left(\frac{\|\bm{a}\|_\alpha u}{a_{\beta_i}}\right).
\end{equation*}
Thus
\begin{equation*}
    \nu_{\bm{X}}(\mathcal{C}_{\beta,u}) = \sum_{j:\frac{\bm{a}_j}{\|\bm{a}_j\|_\alpha}\in C_\beta} \|\bm{a}_j\|_\alpha^\alpha \int_{\max_{i}\left(\frac{\|\bm{a}_j\|_\alpha u}{a_{\beta_i,j}}\right)}^\infty \alpha r^{-\alpha-1}\,\dee r 
    =  \sum_{j:\frac{\bm{a}_j}{\|\bm{a}_j\|_\alpha}\in C_\beta} \min_{i=1,\ldots,s} \left(\frac{a_{\beta_i,j}}{u}\right)^\alpha.
\end{equation*}
For $\alpha=1$ we can establish the upper bound $\hat{\mathbb{P}}(\bm{X}\in\mathcal{C}_{\beta,u})\leq H_{\bm{X}}(C_{\beta})/(su)$ with equality attained if and only if the angular measure places all of its mass on $C_\beta$ at the centroid $\bm{e}(\beta)/|\beta|$, where $\bm{e}(\beta):=(\mathbbm{1}\{i\in\beta\}:i=1,\ldots,d)\in\{0,1\}^d$. This bound represents the limiting probability of the angle lying in the relevant subspace multiplied by the survivor function of a $\mathrm{Pareto}(\alpha=1)$ random variable evaluated at the effective radial threshold $\|u\bm{1}_s\|_1=su$.

\subsubsection{Existing approaches to inference for max-linear models}\label{sec:param-estimation-max-linear}

Suppose $\bm{X}\sim\mathrm{MaxLinear}(A,\alpha)$, where $\alpha$ is known and $A$ must be estimated from a sample $\{\bm{x}_t=(r_t,\bm{\theta}_t):t=1,\ldots,n\}$ of $\bm{X}=(R,\bm{\Theta})$. For $j\in\{1,\ldots,n\}$, let $r_{(j)}$ denote the $j$th upper order statistic of $\{r_1,\ldots,r_n\}$ and let $\bm{x}_{(j)}, \bm{\theta}_{(j)}$ denote the corresponding observation and angular component, respectively.

One estimate of $A$ is motivated by \eqref{eq:max-linear-H}, which says its normalised columns represent the set of realisable extremal angles \citep{cooley2019}.

\begin{definition}\label{def:empirical-A}
    The empirical estimate of $A$ based on $\bm{x}_1,\ldots,\bm{x}_n$ is $\hat{A}=(\hat{\bm{a}}_{1},\ldots,\hat{\bm{a}}_{k})\in\R_+^{d\times k}$, where $1\leq k \leq n$ and $\hat{\bm{a}}_{j} = (d/k)^{1/\alpha} \bm{\theta}_{(j)}$ for $j=1,\ldots,k$.
\end{definition}

The quantity $k$ is the customary tuning parameter that represents the number of 'extreme' observations with norm not less than the implied radial threshold $r_{(k)}$. The associated angular measure (for any $\alpha$) and TPDM (for $\alpha=2$) are given by
\begin{align}
\hat{H}_{\bm{X}}(\cdot) 
&:= H_{\hat{A}\times_{\max}\bm{Z}}(\cdot)
= \frac{d}{k}\sum_{t=1}^n \mathbbm{1}\{\bm{\theta}_t \in\cdot, r_t \geq r_{(k)}\} \label{eq:empirical-H} \\
\hat{\sigma}_{\bm{X}_{ij}} 
&:= \sigma_{\hat{A}\times_{\max}\bm{Z},ij}
= \frac{d}{k}\sum_{t=1}^n \theta_{ti}\theta_{tj}\mathbbm{1}\{r_t \geq r_{(k)}\}
= [\hat{A}\hat{A}^T]_{ij}. \label{eq:empirical-tpdm}
\end{align}
These are the empirical angular measure \citep{einmahl2009} and empirical TPDM \citep{cooley2019}, respectively. In view of \eqref{eq:empirical-tpdm}, alternative estimates of $A$ can be obtained by CP decomposition of the empirical TPDM. 

\begin{definition}\label{def:cp-A}
    CP-factors of $\hat{\Sigma}_{\bm{X}}=(\hat{\sigma}_{\bm{X}_{ij}})$ are called CP-estimates of $A$, denoted $\tilde{A}$.
\end{definition}

An estimate $\tilde{A}$ induces a different angular measure $\tilde{H}_{\bm{X}}:=H_{\tilde{A}\times_{\max}\bm{Z}}$ but, by construction, the empirical and CP models share identical tail pairwise dependencies, since $\tilde{\Sigma}_{\bm{X}}:=\Sigma_{\tilde{A}\times_{\max}\bm{Z}}=\tilde{A}\tilde{A}^T=\hat{\Sigma}_{\bm{X}}$. Note that $\tilde{A}$ implicitly depends on the same tuning parameter $k$ as $\hat{A}$ via the empirical TPDM. All CP-estimates in this paper are obtained using the decomposition algorithm of \cite{kiriliouk2022}, which efficiently factorises moderate- to high-dimensional TPDMs. Their algorithm takes as input a strictly positive TPDM and a permutation $(i_1,\ldots,i_d)$ of $\mathbb{V}_d$ and (for some permutations) returns a square CP-factor $\tilde{A}\in\R_+^{d\times d}$ whose columns satisfy $\tilde{\bm{a}}_{j} / \|\tilde{\bm{a}}_{j}\|_2 \in C_{\mathbb{V}_d\setminus\{i_l \,:\, l < j\}}$ for $j=1,\ldots,d$.

\subsubsection{Inference for max-linear models based on sparse projections}

A limitation of $\hat{A}$ and $\tilde{A}$ is that they fail to capture the true extremal directions of $\bm{X}$. In the case of $\hat{A}$ this is because the angles $\bm{\Theta}=\bm{X}/\|\bm{X}\|$ lie in the simplex interior almost surely, meaning $\hat{\bm{a}}_1,\ldots,\hat{\bm{a}}_k\in\mathbb{V}_d$ and $\hat{\mathbb{P}}(\hat{A}\times_{\max}\bm{Z}\in\mathcal{C}_{\beta,\bm{u}})=0$ for any $\beta\neq\mathbb{V}_d$. The $d$ extremal directions of $\tilde{A}\times_{\max}\bm{Z}$ are fully determined by the input path $(i_1,\ldots,i_d)$ and need not bear any resemblance to the extremal directions suggested by the data. To address this gap, we propose augmenting the empirical estimate with an alternative notion of angle based on Euclidean projections onto the $L_1$-simplex \citep{meyer2023}. 

\begin{definition}
    The Euclidean projection onto the $L_1$-simplex is defined by
    \begin{equation*}
    \pi:\R_+^d\to\mathbb{S}_+^{d-1}, \qquad \pi(\bm{v})=\argmin_{\bm{w}\in\mathbb{S}_+^{d-1}}\|\bm{w} - \bm{v}\|_2^2.
\end{equation*}
\end{definition}

This projection is useful because $\pi(\bm{v})$ may lie on a boundary of the simplex even if $\bm{v}/\|\bm{v}\|_1$ lies in its interior. Assume now that $\alpha=1$.

\begin{definition}\label{def:sparse-empirical-A}
    The \textit{sparse empirical estimate} of $A$ based on $\bm{x}_1,\ldots,\bm{x}_n$ is $\hat{A}^\star=(\hat{\bm{a}}_{1}^\star,\ldots,\hat{\bm{a}}_{k}^\star)\in\R_+^{d\times k}$, where $1\leq k< n$ and $\hat{\bm{a}}_j^\star = (d/k)\pi(\bm{x}_{(j)}/r_{(k+1)})$ for $j=1,\ldots,k$.
\end{definition}
The corresponding angular measure
\begin{equation*}
    \hat{H}_{\bm{X}}^\star(\cdot)
    :=H_{\hat{A}^\star\times_{\max}\bm{Z}}(\cdot) 
    = \frac{d}{k} \sum_{j=1}^k \mathbbm{1}\{\pi\left(\bm{x}_{(j)}/r_{(k+1)}\right) \in\cdot\}
\end{equation*}
spreads mass across the subspaces $C_\beta\subseteq\mathbb{S}_+^{d-1}$ on which the projected data lie and $\hat{\mathbb{P}}(\hat{A}^\star\times_{\max}\bm{Z}\in\mathcal{C}_{\beta,\bm{u}})\neq 0$ for all corresponding $\beta$. A full study of the theoretical properties of $\hat{A}^\star$ has not been conducted. Having introduced our estimator and all the requisite theory, we are ready to present our methods for the multivariate challenges.

\subsection{Challenge 3}

\subsubsection{Data}

Challenge 3 considers a trivariate random vector $\bm{Y}=(Y_1,Y_2,Y_3)$ on standard Gumbel margins, i.e. $\mathbb{P}(Y_i<y)=G(y):=\exp(-\exp(-y))$ for $y\in\R$ and $i=1,2,3$, and entails estimating
\begin{equation*}
    p_1 := \mathbb{P}(Y_1 > y, Y_2 > y, Y_3 > y), \qquad p_2:= \mathbb{P}(Y_1 > v, Y_2 > v, Y_3 < m)
\end{equation*}
where $y=6$, $v=7$, and $m=G^{-1}(1/2)=-\log(\log(2))$. The data comprise $n=21,000$ independent observations $\{\bm{y}_t=(y_{t1},y_{t2},y_{t3}):t=1,\ldots,n\}$ of $\bm{Y}$. Additional covariate information is available but not leveraged by our method.

\subsubsection{Methodology}

Letting $\bm{X}=(X_1,X_2,X_3)$ denote the random vector obtained by transforming $\bm{Y}$ to Fréchet margins with shape parameter $1$, i.e. $X_i=\Phi^{-1}(G(Y_i))=\exp(Y_i)\sim\Phi_1$ for $i=1,2,3$, we rewrite the above probabilities as
\begin{equation}\label{c3-probabilities}
    p_1 = \mathbb{P}(X_1 > e^y, X_2 > e^y, X_3 > e^y), \qquad p_2= \mathbb{P}(X_1 > e^v, X_2 > e^v, X_3 < e^m)
\end{equation}
The new threshold values are $\exp(y)\approx 403\approx\Phi_1^{-1}(0.998)$, $\exp(v)\approx 1097\approx\Phi_1^{-1}(0.999)$ and $\exp(m)=\Phi_1^{-1}(1/2)\approx 1.4$. Our approach is to model $\bm{X}$ as a max-linear random vector, estimate its noise coefficient matrix using the sparse empirical estimate $\hat{A}^\star$ based on the $k$ largest observations, and compute estimates of $p_1$ and $p_2$ using \eqref{eq:prob-approx-formula}, that is
\begin{equation}\label{c3-probability-estimates}
    \hat{p}_1 = \hat{\mathbb{P}}(\hat{A}^\star\times_{\max}\bm{Z}\in\mathcal{C}_{\mathbb{V}_3,\exp(y)}), \qquad \hat{p}_2 = \hat{\mathbb{P}}(\hat{A}^\star\times_{\max}\bm{Z}\in\mathcal{C}_{\{1,2\},\exp(v)}).
\end{equation}
Inference is based on the transformed data $\{\bm{x}_t=(x_{t1},x_{t2},x_{t3}):t=1,\ldots,n\}$, where $x_{ti}:=\exp(y_{ti})$ for $t=1,\ldots,n$ and $i=1,2,3$. 

\subsubsection{Results}

We now apply the method to the (pre-processed) Challenge 3 data. The results presented are based on selecting $k=500$, that is $k/n\approx 2.5\%$. The sensitivity of the results to this choice will be analysed later.

\begin{figure}
    \centering
    \includegraphics[width=\linewidth]{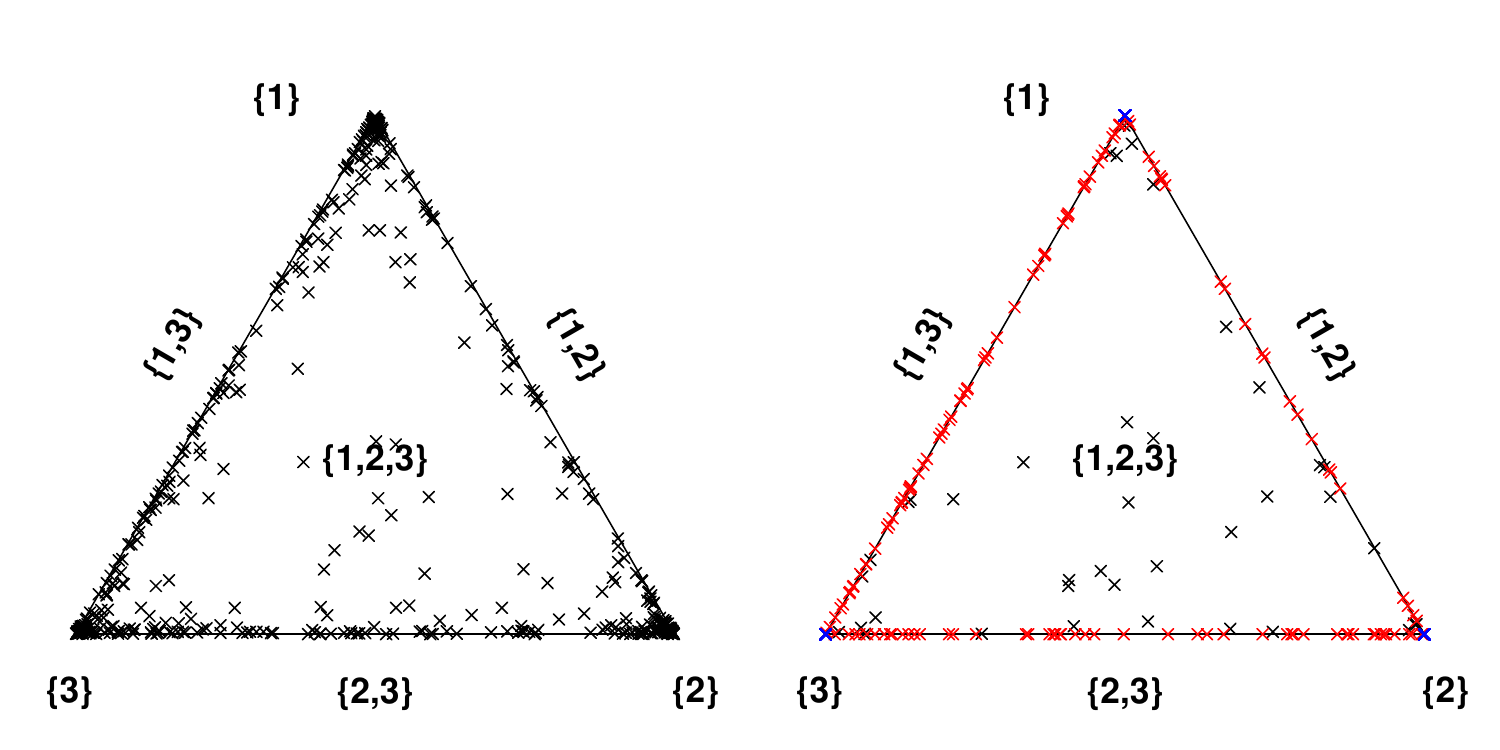}
    \caption{Ternary plots of the angles $\bm{\theta}_{(j)}\in\mathbb{S}_+^2$ (left) and Euclidean projections $\pi(\bm{x}_{(j)}/r_{(k+1})\in\mathbb{S}_+^2$ (right) associated with the $k=500$ largest observations in norm of the (pre-processed) Challenge 3 data. Colours represent whether the point lies in the interior (black), on an edge (red) or on a vertex (blue).}
    \label{fig:ternary}
\end{figure}

The ternary plots in \autoref{fig:ternary} depicts the angles associated with the $k$ largest observations in norm, i.e. exceedances of the radial threshold $r_{(k+1)}\approx 138.77$. The angles in the left-hand plot are the self-normalised vectors  $\bm{\theta}_{(1)},\ldots,\bm{\theta}_{(k)}$,  There are points lying in the interior of the ternary plot as well as in the neighbourhood of each of three edges and three vertices, which suggests that the angular measure spreads mass across all seven subspaces $\{C_\beta: \beta\in\mathcal{P}_3^\star\}$. However, we reiterate that none of the points lie exactly on the boundary. The right-hand plot shows the projected vectors $\{\pi\left(\bm{x}_{t}/r_{(k+1)}\right) : r_t \geq r_{(k)}\}$. The points' colours indicate whether they lie in the interior ($C_{\mathbb{V}_3}$, black, 40 points), on an edge ($C_{\beta}$ with $|\beta|=2$, red, 139 points), or on a vertex ($C_{\beta}$ with $|\beta|=1$, blue, 321 points) of the simplex. Only the 40 vectors in $C_{\mathbb{V}_3}$ and 23 vectors in $C_{\{1,2\}}$ will affect the estimates of $p_1$ and $p_2$.

\begin{figure*}
    \centering
    \includegraphics[width=\linewidth]{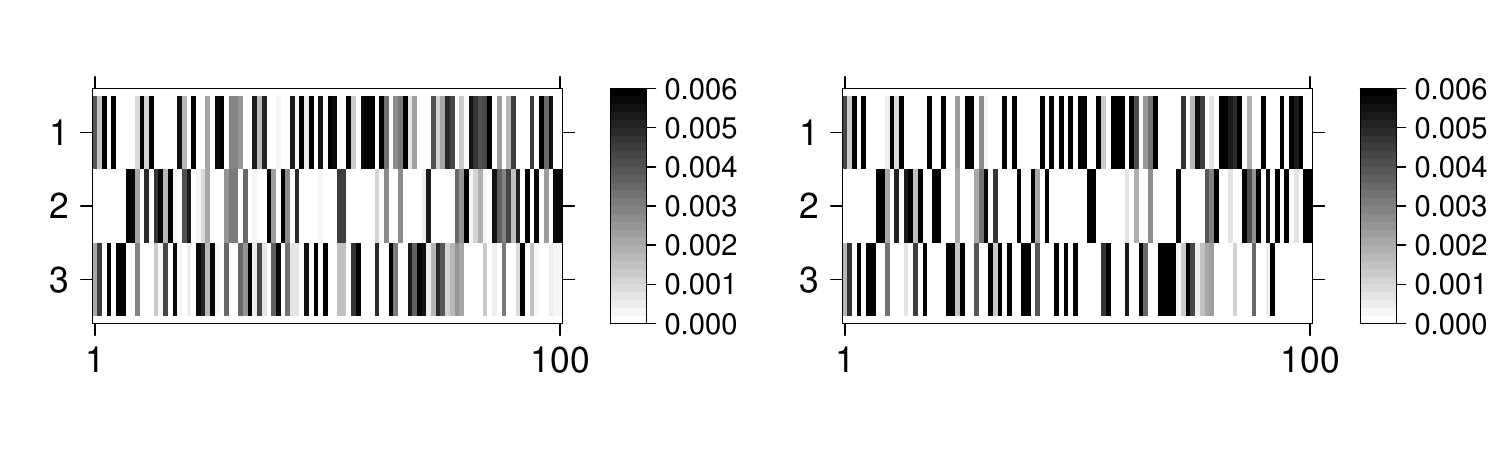}
    \caption{The first 100 columns of $\hat{A}$ (left) and $\hat{A}^\star$ (right) for the Challenge 3 data. The colour intensity of each cell represents the magnitude of the corresponding matrix entry.}
    \label{fig:A-estimates}
\end{figure*}

Next, we collate the projected vectors to form the $3\times 500$ matrix $\hat{A}^\star$. The first 100 columns of $\hat{A}^\star$ are shown on the right-hand plot in \autoref{fig:A-estimates}; the leading columns of $\hat{A}$ are included on the left for comparison. The matrix $\hat{A}^\star$ is sparser than $\hat{A}$, i.e. $\|\hat{\bm{a}}_j^\star\|_0\leq \|\hat{\bm{a}}_j\|_0$ for all $j=1,\ldots,k$, and typically this inequality is strict. 
Unlike $\hat{A}$, the sparse matrix $\hat{A}^\star$ has duplicate columns; merging and appropriately re-weighting these columns would yield a compressed estimate $\hat{A}_{\mathrm{comp}}^\star$ with only $q_{\mathrm{comp}}=40+139+3=182$ columns, but the random vectors $\hat{A}^\star\times_{\max}\bm{Z}$ and $\hat{A}_{\mathrm{comp}}^\star\times_{\max}\bm{Z}$ have the same angular measure. 

\begin{figure}
    \centering
    \includegraphics[width=\textwidth]{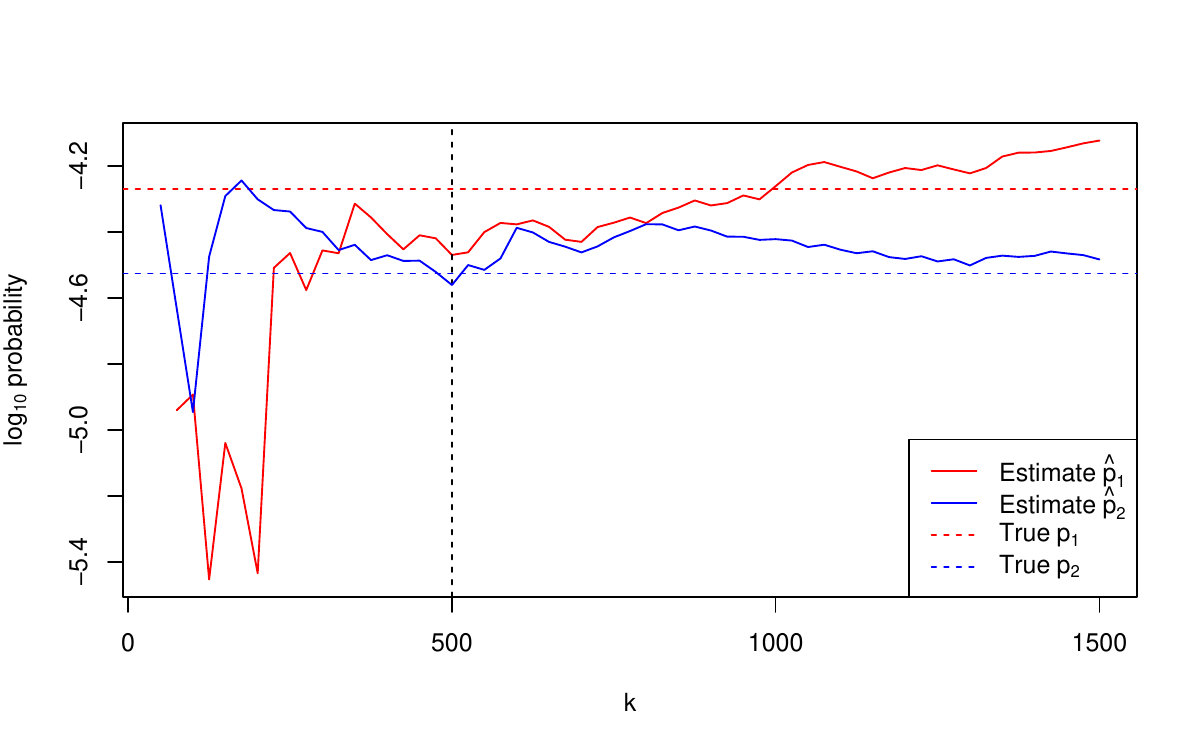}
    \caption{Estimates of $p_1$ and $p_2$ from Challenge 3 from repeated application of our methodology while varying the number of extreme observations, $k$, compared against the true values (horizontal dashed lines). Our submitted values were based on $k=500$ (vertical dashed line).}
    \label{fig:k-sensitivity}
\end{figure}

Substituting $\hat{A}^\star$ or $\hat{A}_{\mathrm{comp}}^\star$ into \eqref{c3-probability-estimates} yields the final point estimates $\hat{p}_1=3.36\times 10^{-5}$ and $\hat{p}_2=2.76\times 10^{-5}$, to three significant figures. We are pleased to find that our estimates are very close to the true values given in \cite{Rohr23}. \autoref{fig:k-sensitivity} shows that the estimates are fairly stable within the range of sensible choices of $k$, say, $250\leq k\leq 750$. This range corresponds to a sampling fraction of roughly $1\%\leq k/n \leq 5\%$. When the effective sample size becomes prohibitively small ($k<250$) the estimates are highly variable.

\subsection{Challenge 4}

\subsubsection{Data}

Challenge 4 regards a $d=50$ dimensional random vector $\bm{Y}$ on standard Gumbel margins. The components of $\bm{Y}$ are random variables $Y_{i,j}$ for $i=1,\ldots,25$ and $j=1,2$, representing the value of an environmental variable at the $i$th site in the administrative area of government area $j$. The estimands are the probabilities
\begin{align*}
    p_1 &:= \mathbb{P}(Y_{i,j} > G^{-1}(1-\phi_j) : i=1,\ldots,25, \, j=1,2), \\
    p_2 &:= \mathbb{P}(Y_{i,j} > G^{-1}(1-\phi_1) : i=1,\ldots,25, \, j=1,2),
\end{align*}
where $\phi_1=1/300$ and $\phi_2=12/300$, based on $n=10,000$ independent observations $\{\bm{y}_t=(y_{t,i,j}:i=1,\ldots,25,\,j=1,2):t=1,\ldots,n\}$ of $\bm{Y}$.

\subsubsection{Methodology}

Let $\bm{X}=(X_1,\ldots,X_d)$ denote the random vector obtained from $\bm{Y}$ by re-indexing its variables and transforming to Fréchet margins with shape parameter $\alpha=2$, i.e.
\begin{equation*}
    X_{i+25(j-1)}:= \Phi_2^{-1}(G(Y_{i,j})) = \exp(Y_{i,j}/2)\sim\Phi_2, \qquad (i=1,\ldots,25,\,j=1,2).
\end{equation*}
The choice of $\alpha=2$ will be justified later. The data are transformed in an identical way yielding $\{\bm{x}_t=(x_{t1},\ldots,x_{td}):t=1,\ldots,n\}$, where $x_{t,i+25(j-1)}:=\exp(y_{t,i,j}/2)$ for $t=1,\ldots,n$, $i=1,\ldots,25$ and $j=1,2$. The above probabilities can be equivalently expressed as
\begin{align*}
    p_1=\mathbb{P}(\bm{X}\in \mathcal{C}_{\mathbb{V}_d,\bm{u}_1}), \qquad & [\bm{u}_1]_i = 
    \begin{cases}
        \Phi_2^{-1}(1-\phi_1), & \text{if } 1 \leq i \leq 25 \\
        \Phi_2^{-1}(1-\phi_2), & \text{if } 26 \leq i \leq 50
    \end{cases}, \\
    p_2=\mathbb{P}(\bm{X}\in \mathcal{C}_{\mathbb{V}_d,\bm{u}_2}), \qquad &\bm{u}_2 = \Phi_2^{-1}(1-\phi_1)\bm{1}_{50}.
\end{align*}

At a high level, our method is the same as for Challenge 3: we model $\bm{X}$ using the max-linear model, estimate the parameter $A$, and compute approximate estimates of $p_1$ and $p_2$ based on \eqref{eq:prob-approx-formula}. However, our exploratory analysis revealed that not all components of $\bm{X}$ are asymptotically dependent, implying that $H_{\bm{X}}(\mathbb{V}_d)=0$. The angular measures that arise from the empirical and CP-estimates of $A$ never satisfy this property; their use would result in a misspecified model. The sparse empirical estimate can satisfy $\hat{H}^\star_{\bm{X}}(\mathbb{V}_d)=0$, but then we face the issue that $\hat{\mathbb{P}}(\hat{A}^\star\times_{\max}\bm{Z}\in\mathcal{C}_{\mathbb{V}_d,\bm{u}})=0$ for any $\bm{u}>\bm{0}$. However, for a finite threshold vector $\bm{u}$ the probability $\mathbb{P}(\bm{X}\in\mathcal{C}_{\mathbb{V}_d,u})$ may be non-zero. We deal with this discrepancy by identifying clusters of asymptotically dependent variables prior to the model fitting step. To this end, we impose an additional assumption that the marginal variables can be partitioned such that asymptotic independence is present between clusters but not within them.

\begin{assumption}\label{assump:fomichov}
There exists $2\leq K \leq d$ and a partition $\beta_1,\ldots,\beta_K$ of $\mathbb{V}_d$ such that the angular measure is supported on the closed subspaces $\bar{C}_{\beta_1},\ldots,\bar{C}_{\beta_K}\subset\mathbb{S}_+^{d-1}$, where $\bar{C}_\beta := \{C_{\beta'}:\beta'\subseteq\beta\}$ for any $\beta\in\mathcal{P}_d^\star$. That is, $H_{\bm{X}}(\cup_{l=1}^K \bar{C}_{\beta_l})=m$.
\end{assumption}
This scenario has been considered before, cf. Assumption 1 in \cite{fomichov2023}. If $\bm{X}$ is max-linear with parameter $A=(\bm{a}_{1},\ldots,\bm{a}_{q})\in\R_{+}^{d\times q}$, then Assumption \ref{assump:fomichov} can be equivalently restated as follows.

\begin{assumption}\label{assump:fomichov-max-linear}
    There exist permutations $\pi:\mathbb{V}_d\to\mathbb{V}_d$ and $\phi:\mathbb{V}_q\to\mathbb{V}_q$ such that $\bm{X}_{\pi} := (X_{\pi(1)},\ldots, X_{\pi(d)}) \sim\mathrm{MaxLinear}(A_\phi)$, where $A_\phi := \left( \bm{a}_{\phi(1)},\ldots,\bm{a}_{\phi(q)}\right)=\mathrm{diag}(A_\phi^{(1)},\ldots,A_\phi^{(K)})\in\R_{+}^{d\times q}$, i.e. $A_\phi$ is block-diagonal with $2\leq K\leq d$ blocks. For $l=1,\ldots,K$, the $l$th block matrix $A_\phi^{(l)}$ has $d_l=|\beta_l|$ rows, $1\leq q_l < q$ columns, and is such that the $q_l\times q_l$ matrix $A_\phi^{(l)}(A_\phi^{(l)})^T$ has strictly positive entries. The blocks' dimensions satisfy $\sum_{l=1}^K d_l = d$ and $\sum_{l=1}^K q_l = q$.
\end{assumption}

Under Assumption \ref{assump:fomichov-max-linear}, $\bm{X}$ divides into random sub-vectors $\bm{X}^{(1)},\ldots,\bm{X}^{(K)}$, where $\bm{X}^{(l)}:=(X_j:j\in\beta_l)\sim\mathrm{MaxLinear}(A_\phi^{(l)})$ for $l=1,\ldots,K$. (Henceforth, assume that the columns of $A$ are correctly ordered, so that $\phi$ is the identity permutation and $A=A_\phi$.) The clustering assumption is simplistic and cannot be expected to hold in general applications, but it reflects what we suspect to be the true dependence structure for Challenge 4. Moreover, it is dimension reducing because the original $d$-dimensional problem is transformed to a set of $K$ independent problems with dimensions $d_1,\ldots,d_K<d$. This ameliorates the curse of dimensionality to some degree.

In general, the full joint exceedance event can be decomposed into concurrent joint exceedances in all $K$ clusters as $\{\bm{X}\in\mathcal{C}_{\mathbb{V}_d,u}\}=\cap_{l=1}^K \{\bm{X}\in\mathcal{C}_{\mathbb{V}_{d_l},\bm{u}^{(l)}}\}$, where each $\bm{u}^{(l)}$ is a sub-vector of $\bm{u}$ defined analogously to $\bm{X}^{(l)}$. Since we consider variables in different clusters to be asymptotically independent, we assume that for large, finite thresholds, joint exceedances in each cluster are approximately independent events, so that 
\begin{equation*}
    \mathbb{P}(\bm{X}\in\mathcal{C}_{\mathbb{V}_d,\bm{u}}) 
    = \mathbb{P}\left(\bigcap_{l=1}^K \{\bm{X}^{(l)}\in\mathcal{C}_{\mathbb{V}_{d_l},\bm{u}^{(l)}}\}\right) 
    \approx \prod_{l=1}^K \mathbb{P}(\bm{X}^{(l)}\in\mathcal{C}_{\mathbb{V}_{d_l},\bm{u}^{(l)}}).
\end{equation*}
Assuming $\bm{X}^{(l)}\sim\mathrm{MaxLinear}(A^{(l)})$ for $l=1,\ldots,K$, each term in the product can be estimated using~\eqref{eq:prob-approx-formula}, so that 
\begin{equation}\label{c4-probability-estimates}
    \mathbb{P}(\bm{X}\in\mathcal{C}_{\mathbb{V}_d,\bm{u}}) \approx \prod_{l=1}^K \hat{\mathbb{P}}(A^{(l)}\times_{\max}\bm{Z}\in\mathcal{C}_{\mathbb{V}_{d_l},\bm{u}^{(l)}}).
\end{equation}
The final step is to replace $A^{(1)},\ldots,A^{(K)}$ with suitable estimates. Any of the empirical, sparse empirical, or CP-estimates are viable options.
We opted to use CP estimation for two reasons: (i) it is rooted in the TPDM, which is geared towards high-dimensional settings, and (ii) it is non-unique, enabling us to compute many parameter estimates and, correspondingly, many probability estimates. The variation in these sets of estimates reflects the model uncertainty that arises from summarising dependence via the TPDM and thereby overlooking higher-order dependencies between components. The use of CP-estimates justifies our choosing $\alpha=2$.

\subsubsection{Results}

We now present our results for Challenge 4. First, the variables $X_1,\ldots,X_d$ are partitioned into $K$ groups based on asymptotic (in)dependence using the clustering algorithm of \cite{bernard2013}. This entails constructing a distance matrix $\mathcal{D}=(\hat{d}_{ij})$, where $\hat{d}_{ij}$ denotes a non-parametric estimate of the F-madogram distance between variables $X_i$ and $X_j$. The distance metric is connected to the strength of extremal dependence between $X_i$ and $X_j$, with $\hat{d}_{ij}\approx 0$ implying strong asymptotic dependence and $\hat{d}_{ij}=1/6$ in the case of asymptotic independence. The partition around medoids (PAM) clustering algorithm \citep{kaufman1990} returns a partition $\beta_1,\ldots,\beta_{K}$ of $\mathbb{V}_d$ based on $\mathcal{D}$. The number of clusters $K$ is a prespecified tuning parameter; we identify $K=5$ clusters whose sizes are given in \autoref{tab:cluster-summary}. Defining cluster membership variables $\mathcal{M}_1,\ldots,\mathcal{M}_d\in\{1,\ldots,K\}$ by $\mathcal{M}_i=l\iff i\in\beta_l$ for $i=1,\ldots,d$, we find that $\max\{\hat{d}_{ij}: \mathcal{M}_i=\mathcal{M}_j\} = 0.113 < 1/6$ and $\min\{\hat{d}_{ij}:\mathcal{M}_i\neq \mathcal{M}_j\} = 0.164 \approx 1/6$. These summary statistics are consistent with Assumption \ref{assump:fomichov}.

\begin{figure}
    \centering
    \includegraphics[width=\linewidth]{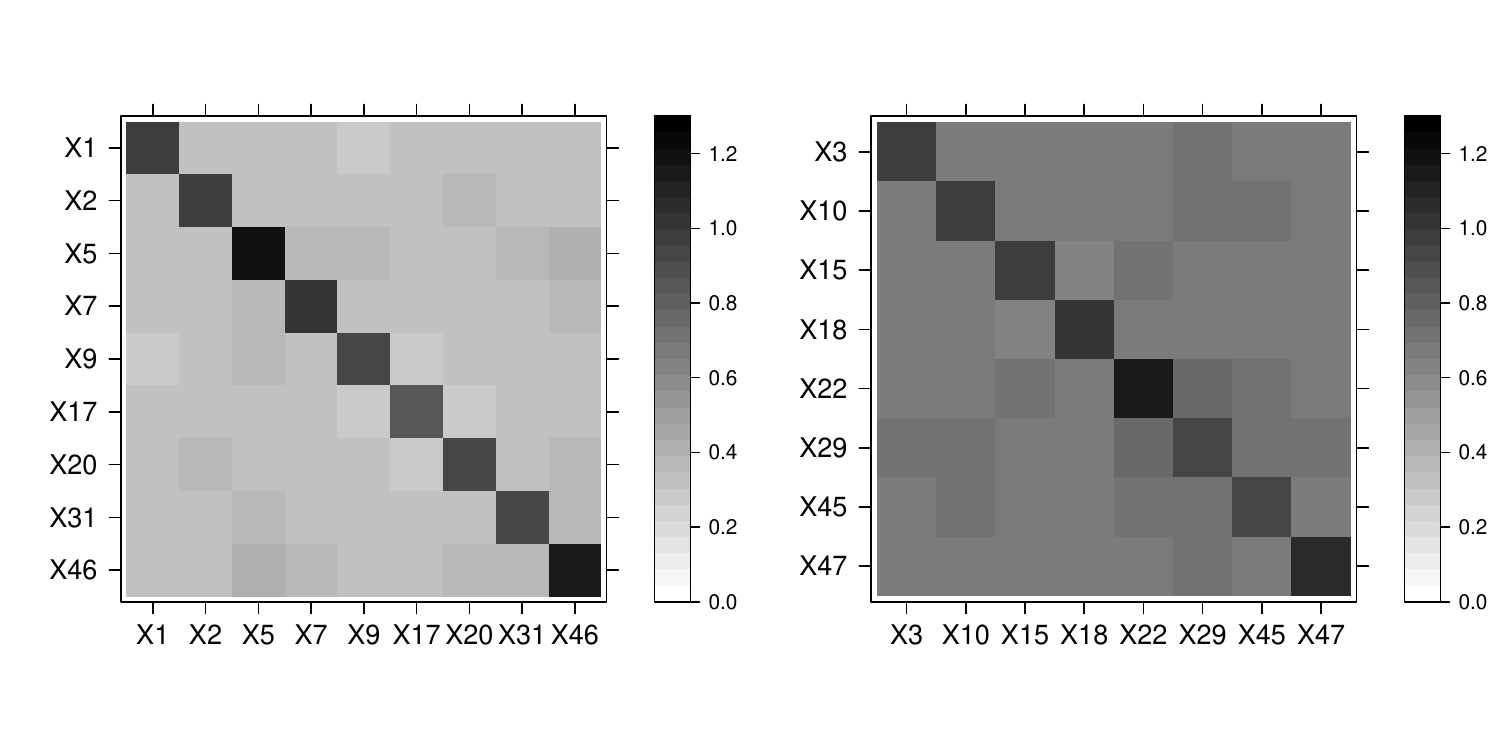}
    \caption{The estimated TPDMs for the first two clusters, $\hat{\Sigma}_{\bm{X}^{(1)}}$ (left) and $\hat{\Sigma}_{\bm{X}^{(2)}}$ (right), based on the $k=250$ most extreme observations in norm.}
    \label{fig:empirical-tpdm}
\end{figure}

\begin{table}[t]
\caption{Summary statistics for the Challenge 4 clusters and the estimated TPDMs.}
\label{tab:cluster-summary}
\centering
\begin{tabular}{@{}lllllll@{}}
\toprule
\multirow{2}{*}{Cluster, $l$} & \multirow{2}{*}{Size, $d_l$} & \multirow{2}{*}{U1 sites} & \multirow{2}{*}{U2 sites} & \multicolumn{3}{c}{$\{\hat{\sigma}_{\bm{X}^{(l)}_{ij}} : i\neq j\}$} \\ \cmidrule(l){5-7} 
 &  &  &  & Min. & Median & Max. \\ \midrule
1 & 9 & 7 & 2 & 0.30 & 0.33 & 0.40 \\
2 & 8 & 5 & 3 & 0.64 & 0.68 & 0.74 \\
3 & 8 & 3 & 5 & 0.62 & 0.67 & 0.74 \\
4 & 12 & 7 & 5 & 0.27 & 0.33 & 0.39 \\
5 & 13 & 3 & 10 & 0.43 & 0.50 & 0.58 \\ \bottomrule
\end{tabular}
\end{table}
Next, we estimate the TPDMs of $\bm{X}^{(1)},\ldots,\bm{X}^{(K)}$. This step is a prerequisite for obtaining the CP-estimates of $A^{(1)},\ldots,A^{(K)}$. For $l=1,\ldots,K$ and $t=1,\ldots,n$, define the observational sub-vector $\bm{x}_t^{(l)}=(x_{ti}:i\in\beta_l)$ and its radial and angular components $r_t^{(l)}=\|\bm{x}_t^{(l)}\|_2$, $\bm{\theta}_t^{(l)}=\bm{x}_t^{(l)}/\|\bm{x}_t^{(l)}\|_2$, respectively. Let $\bm{x}_{(j)}^{(l)}$,~$r_{(j)}^{(l)}$,~$\bm{\theta}_{(j)}^{(l)}$ denote the vector, radius, and angle associated with the $j$th largest observation in norm among $\bm{x}_1^{(l)},\ldots,\bm{x}_n^{(l)}$. Choose a tuning parameter $1\leq k_l \leq n$ representing the number of extreme observations on which to base the empirical TPDM of cluster $l$, which is computed via the empirical estimate of $A^{(l)}$ as
\begin{equation*}
    \hat{A}^{(l)} = \left( \frac{d_l}{k_l}\bm{\theta}_{(1)}^{(l)} ,\ldots,\frac{d_l}{k_l}\bm{\theta}_{(k_l)}^{(l)}\right),\qquad \hat{\Sigma}_{\bm{X}^{(l)}}=\hat{A}^{(l)}(\hat{A}^{(l)})^T.
\end{equation*}
We set $k_1=\ldots=k_K=:k=250$, corresponding to a sampling fraction of $k/n=2.5\%$ for each cluster. The empirical TPDM estimates for the first two clusters are displayed in \autoref{fig:empirical-tpdm} and summary statistics for all five empirical TPDMs are listed in \autoref{tab:cluster-summary}. Asymptotic dependence is strongest in clusters 2 and 3 and weakest in cluster 1 and 4.

\begin{figure}
    \centering
    \includegraphics[width=\linewidth]{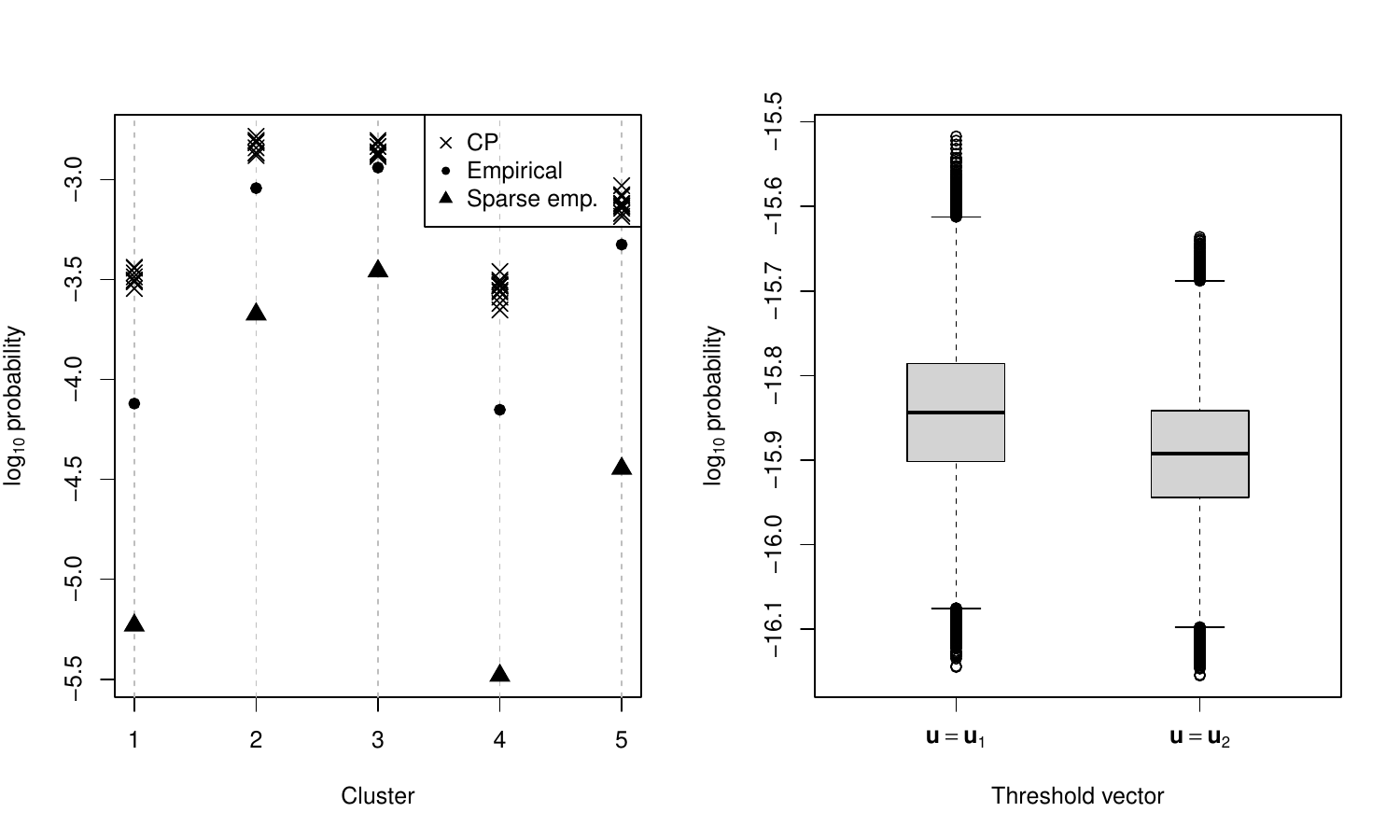}
    \caption{Left: the (distinct) estimates of $\mathbb{P}(\bm{X}^{(l)}\in\mathcal{C}_{\mathbb{V}_{d_l},\bm{u}_1^{(l)}})$ based on CP estimates (crosses), empirical estimates (circles) and sparse empirical estimates (triangles) of $A^{(l)}$. Right: the distributions of $\tilde{P}_1$ and $\tilde{P}_2$.}
    \label{fig:prob-estimates}
\end{figure}

By repeated application of the CP-decomposition algorithm of \cite{kiriliouk2022} with randomly chosen inputs $(i_1,\ldots,i_{d_l})$, we obtain $N_{\mathrm{cp}}=50$ CP-estimates of each $A^{(l)}$, denoted by $\tilde{A}^{(l)}_1,\ldots,\tilde{A}^{(l)}_{N_{\mathrm{cp}}}$. Note that among $\tilde{A}^{(l)}_1,\ldots,\tilde{A}^{(l)}_{N_{\mathrm{cp}}}$ there are at most $d_l$ distinct leading columns, because there are only $d_l$ different ways to initialise the (deterministic) algorithm. Since $\hat{\mathbb{P}}(\tilde{A}\times \bm{Z}\in\mathcal{C}_{\mathbb{V}_{d},\bm{u}})$ is fully determined by $\tilde{\bm{a}}_1$, the set $\{\hat{\mathbb{P}}(\tilde{A}_r^{(l)}\times \bm{Z}\in\mathcal{C}_{\mathbb{V}_{d_l},\bm{u}^{(l)}}) : r=1,\ldots,N_{\mathrm{cp}}\}$ contains at most $d_l$ distinct values. These values are plotted as crosses in \autoref{fig:prob-estimates} for $\bm{u}=\bm{u}_1$. Clusters 2 and 3 are deemed most likely to experience a joint extreme event, because they contain a small number of variables ($d_2=d_3=8$) and exhibit strong dependence (\autoref{fig:empirical-tpdm}, right). The lower risk in clusters 1 and 4 can be attributed to weak dependence (\autoref{fig:empirical-tpdm}, left) and being primarily composed of sites in area U1 (\autoref{tab:cluster-summary}), where the thresholds are higher.

Substituting each CP-estimate into \eqref{c4-probability-estimates} and enumerating over all possible combinations, we produce sets of estimates of $p_i$ given by
\begin{multline*}
    \tilde{P}_i:=\{ \hat{\mathbb{P}}(\tilde{A}^{(1)}_{r_1}\times_{\max}\bm{Z}\in\mathcal{C}_{\mathbb{V}_{d_1},\bm{u}_i^{(1)}}) \times \ldots \times \hat{\mathbb{P}}(\tilde{A}^{(K)}_{r_K}\times_{\max}\bm{Z}\in\mathcal{C}_{\mathbb{V}_{d_K},\bm{u}_i^{(K)}}) \\: r_1,\ldots,r_K = 1,\ldots,N_{\mathrm{cp}} \},
\end{multline*}
for $i=1,2$. Each set has size $N_{\mathrm{cp}}^K\approx 3\times 10^8$ (including repeated values) and contains $\prod_{l=1}^K d_l=89,856$ distinct values. The distributions of the estimates are illustrated in \autoref{fig:prob-estimates}. For final point estimates we take the median values $\tilde{p}_1 := \mathrm{median}(\tilde{P}_1)=1.4\times 10^{-16}$ and $\tilde{p}_2 := \mathrm{median}(\tilde{P}_2)= 1.3\times 10^{-16}$, respectively, to two significant figures.  

\subsubsection{Improving performance using sparse empirical estimates}

Unfortunately, it transpires that our method dramatically overestimated the true probabilities. In hindsight, this could have been anticipated in view of the simulation study in Section 5.1 in \cite{kiriliouk2022}, where the authors remark that failure regions of the the type $\mathcal{C}_{\mathbb{V}_d,\bm{u}}$ seem to be poorly summarised by the TPDM. This prompts us to investigate whether using standard/sparse empirical estimates instead of CP estimates would have improved our performance. For the empirical estimates, this simply involves replacing $A^{(l)}$ with the precomputed matrix $\hat{A}^{(l)}$ in \eqref{c4-probability-estimates}. The approach for sparse empirical estimates follows analogously except that we revert to the $\alpha=1$ setting and transform the data and thresholds accordingly. The resulting joint exceedance probability estimates for each cluster and $\bm{u}=\bm{u}_1$ are shown in \autoref{fig:prob-estimates}. Both empirical estimates lead to lower (which we assume to be better) estimates in all clusters. In fact, the sparse empirical estimates give final point estimates $\hat{p}_1^\star=5.1\times 10^{-23}$ and $\hat{p}_2^\star = 5.0\times 10^{-24}$ for $p_1$ and $p_2$, which are very close to the true values $p_1=8.4\times 10^{-23}$ and $p_2=5.4\times 10^{-25}$. Submitting these values would have significantly improved our ranking for this sub-challenge.

\section{Conclusion}\label{conc}

We propose four different methods in order to solve the EVA Data Challenge, methods that allow for the estimation of extreme quantiles and probabilities. In the univariate cases, we estimate the conditional quantile by attempting to cluster the covariates to categorise the extremes. Once clustered, bootstrapping was used in order to estimate the central 50\% confidence intervals. In practice however, we found no clearly defined clusters and that the Gaussian mixture distribution was not able to capture the non-Gaussian distribution of the covariates. This could also be extended by using the covariates when fitting the GPDs. In the second challenge, we were able to use bootstrapping on simulated datasets in order to identify a bias for estimating small values of the GPD shape parameter. This bias arises when fitting a survival function to data using a MCMC method. Once accounted for, we were able to shift our estimate for the return level from an underestimate to a potentially less-severe overestimate. 

For the multivariate challenges, we show that the max-linear model provides a good framework for estimating the probabilities of tail events. We introduce a new approach for estimating the parameter of this model based on sparse simplex projections, and find that it is capable of achieving exceptional performance on both Challenges 3 and 4. Given these results, further research on the sparse empirical estimate and its theoretical properties is warranted. 



\backmatter





\bmhead{Acknowledgements}

Matthew Pawley is supported by a scholarship from the EPSRC Centre for Doctoral Training in Statistical Applied Mathematics at Bath (SAMBa), under the project EP/S022945/1. Henry Elsom is supported by the University Research Studentship Award (URSA) from the University of Bath. The authors are thankful to Christian Rohrbeck, Emma Simpson and Jonathan Tawn of the University of Bath, University College London, and Lancaster University, respectively, for organising the challenges that inspired this work.


\begin{thebibliography}{28}
\ifx \bisbn   \undefined \def \bisbn  #1{ISBN #1}\fi
\ifx \binits  \undefined \def \binits#1{#1}\fi
\ifx \bauthor  \undefined \def \bauthor#1{#1}\fi
\ifx \batitle  \undefined \def \batitle#1{#1}\fi
\ifx \bjtitle  \undefined \def \bjtitle#1{#1}\fi
\ifx \bvolume  \undefined \def \bvolume#1{\textbf{#1}}\fi
\ifx \byear  \undefined \def \byear#1{#1}\fi
\ifx \bissue  \undefined \def \bissue#1{#1}\fi
\ifx \bfpage  \undefined \def \bfpage#1{#1}\fi
\ifx \blpage  \undefined \def \blpage #1{#1}\fi
\ifx \burl  \undefined \def \burl#1{\textsf{#1}}\fi
\ifx \doiurl  \undefined \def \doiurl#1{\url{https://doi.org/#1}}\fi
\ifx \betal  \undefined \def \betal{\textit{et al.}}\fi
\ifx \binstitute  \undefined \def \binstitute#1{#1}\fi
\ifx \binstitutionaled  \undefined \def \binstitutionaled#1{#1}\fi
\ifx \bctitle  \undefined \def \bctitle#1{#1}\fi
\ifx \beditor  \undefined \def \beditor#1{#1}\fi
\ifx \bpublisher  \undefined \def \bpublisher#1{#1}\fi
\ifx \bbtitle  \undefined \def \bbtitle#1{#1}\fi
\ifx \bedition  \undefined \def \bedition#1{#1}\fi
\ifx \bseriesno  \undefined \def \bseriesno#1{#1}\fi
\ifx \blocation  \undefined \def \blocation#1{#1}\fi
\ifx \bsertitle  \undefined \def \bsertitle#1{#1}\fi
\ifx \bsnm \undefined \def \bsnm#1{#1}\fi
\ifx \bsuffix \undefined \def \bsuffix#1{#1}\fi
\ifx \bparticle \undefined \def \bparticle#1{#1}\fi
\ifx \barticle \undefined \def \barticle#1{#1}\fi
\bibcommenthead
\ifx \bconfdate \undefined \def \bconfdate #1{#1}\fi
\ifx \botherref \undefined \def \botherref #1{#1}\fi
\ifx \bchapter \undefined \def \bchapter#1{#1}\fi
\ifx \bbook \undefined \def \bbook#1{#1}\fi
\ifx \bcomment \undefined \def \bcomment#1{#1}\fi
\ifx \oauthor \undefined \def \oauthor#1{#1}\fi
\ifx \citeauthoryear \undefined \def \citeauthoryear#1{#1}\fi
\ifx \endbibitem  \undefined \def \endbibitem {}\fi
\ifx \bconflocation  \undefined \def \bconflocation#1{#1}\fi
\ifx \arxivurl  \undefined \def \arxivurl#1{\textsf{#1}}\fi
\csname PreBibitemsHook\endcsname

\bibitem[\protect\citeauthoryear{Bernard et~al.}{2013}]{bernard2013}
\begin{barticle}
\bauthor{\bsnm{Bernard}, \binits{E.}},
\bauthor{\bsnm{Naveau}, \binits{P.}},
\bauthor{\bsnm{Vrac}, \binits{M.}},
\bauthor{\bsnm{Mestre}, \binits{O.}}:
\batitle{Clustering of maxima: Spatial dependencies among heavy rainfall in france}.
\bjtitle{Journal of Climate}
\bvolume{26}(\bissue{20}),
\bfpage{7929}--\blpage{7937}
(\byear{2013})
\end{barticle}
\endbibitem

\bibitem[\protect\citeauthoryear{Clarkson et~al.}{2023}]{clarkson23}
\begin{barticle}
\bauthor{\bsnm{Clarkson}, \binits{D.}},
\bauthor{\bsnm{Eastoe}, \binits{E.}},
\bauthor{\bsnm{Leeson}, \binits{A.}}:
\batitle{The importance of context in extreme value analysis with application to extreme temperatures in the u.s. and greenland}.
\bjtitle{Journal of the Royal Statistical Society Series C: Applied Statistics}
\bvolume{72}(\bissue{4}),
\bfpage{829}--\blpage{843}
(\byear{2023})
\end{barticle}
\endbibitem

\bibitem[\protect\citeauthoryear{Coles}{2001}]{Coles2001}
\begin{bbook}
\bauthor{\bsnm{Coles}, \binits{S.}}:
\bbtitle{An Introduction to Statistical Modeling of Extreme Values},
\bedition{1st ed.} edn.
\bpublisher{Springer},
\blocation{London}
(\byear{2001})
\end{bbook}
\endbibitem

\bibitem[\protect\citeauthoryear{Coles and Powell}{1996}]{coles1996}
\begin{barticle}
\bauthor{\bsnm{Coles}, \binits{S.G.}},
\bauthor{\bsnm{Powell}, \binits{E.A.}}:
\batitle{Bayesian methods in extreme value modelling: A review and new developments}.
\bjtitle{International Statistical Review / Revue Internationale de Statistique}
\bvolume{64}(\bissue{1}),
\bfpage{119}--\blpage{136}
(\byear{1996})
\end{barticle}
\endbibitem

\bibitem[\protect\citeauthoryear{Cooley and Thibaud}{2019}]{cooley2019}
\begin{barticle}
\bauthor{\bsnm{Cooley}, \binits{D.}},
\bauthor{\bsnm{Thibaud}, \binits{E.}}:
\batitle{Decompositions of dependence for high-dimensional extremes}.
\bjtitle{Biometrika}
\bvolume{106}(\bissue{3}),
\bfpage{587}--\blpage{604}
(\byear{2019})
\end{barticle}
\endbibitem

\bibitem[\protect\citeauthoryear{Davison and Smith}{1990}]{DavSmith1990}
\begin{barticle}
\bauthor{\bsnm{Davison}, \binits{A.C.}},
\bauthor{\bsnm{Smith}, \binits{R.L.}}:
\batitle{Models for exceedences over high thresholds}.
\bjtitle{Journal of the Royal Statistical Society: Series B (Methodological).}
\bvolume{52}(\bissue{3}),
\bfpage{393}--\blpage{425}
(\byear{1990})
\end{barticle}
\endbibitem

\bibitem[\protect\citeauthoryear{Einmahl and Segers}{2009}]{einmahl2009}
\begin{botherref}
\oauthor{\bsnm{Einmahl}, \binits{J.H.J.}},
\oauthor{\bsnm{Segers}, \binits{J.}}:
Maximum empirical likelihood estimation of the spectral measure of an extreme-value distribution.
The Annals of Statistics
\textbf{37}(5B)
(2009)
\end{botherref}
\endbibitem

\bibitem[\protect\citeauthoryear{Fawcett and Walshaw}{2006}]{FawWalsh06}
\begin{barticle}
\bauthor{\bsnm{Fawcett}, \binits{L.}},
\bauthor{\bsnm{Walshaw}, \binits{D.}}:
\batitle{A hierarchical model for extreme wind speeds}.
\bjtitle{Journal of the Royal Statistical Society: Series C (Applied Statistics)}
\bvolume{55}(\bissue{5}),
\bfpage{631}--\blpage{646}
(\byear{2006})
\end{barticle}
\endbibitem

\bibitem[\protect\citeauthoryear{Fomichov and Ivanovs}{2023}]{fomichov2023}
\begin{barticle}
\bauthor{\bsnm{Fomichov}, \binits{V.}},
\bauthor{\bsnm{Ivanovs}, \binits{J.}}:
\batitle{Spherical clustering in detection of groups of concomitant extremes}.
\bjtitle{Biometrika}
\bvolume{110}(\bissue{1}),
\bfpage{135}--\blpage{153}
(\byear{2023})
\end{barticle}
\endbibitem

\bibitem[\protect\citeauthoryear{{Fougères} et~al.}{2013}]{fougères2013}
\begin{barticle}
\bauthor{\bsnm{{Fougères}}, \binits{A.-L.}},
\bauthor{\bsnm{Mercadier}, \binits{C.}},
\bauthor{\bsnm{Nolan}, \binits{J.P.}}:
\batitle{Dense classes of multivariate extreme value distributions}.
\bjtitle{Journal of Multivariate Analysis}
\bvolume{116},
\bfpage{109}--\blpage{129}
(\byear{2013})
\end{barticle}
\endbibitem

\bibitem[\protect\citeauthoryear{Fraley and Raftery}{2002}]{fraley2002}
\begin{barticle}
\bauthor{\bsnm{Fraley}, \binits{C.}},
\bauthor{\bsnm{Raftery}, \binits{A.}}:
\batitle{Model-based clustering, discriminant analysis, and density estimation}.
\bjtitle{J. Amer. Statist. Assoc.}
\bvolume{97},
\bfpage{611}--\blpage{631}
(\byear{2002})
\end{barticle}
\endbibitem

\bibitem[\protect\citeauthoryear{Fraley and Raftery}{2003}]{fraley2003}
\begin{barticle}
\bauthor{\bsnm{Fraley}, \binits{C.}},
\bauthor{\bsnm{Raftery}, \binits{A.}}:
\batitle{Enhanced model-based clustering, density estimation, and discriminant analysis software: Mclust}.
\bjtitle{Journal of Classification}
\bvolume{20},
\bfpage{263}--\blpage{286}
(\byear{2003})
\end{barticle}
\endbibitem

\bibitem[\protect\citeauthoryear{Gilleland}{2020}]{gill2020}
\begin{barticle}
\bauthor{\bsnm{Gilleland}, \binits{E.}}:
\batitle{Bootstrap methods for statistical inference. part ii: Extreme-value analysis}.
\bjtitle{J. Atmos. Oceanic Technol.}
\bvolume{37},
\bfpage{2135}--\blpage{2144}
(\byear{2020})
\end{barticle}
\endbibitem

\bibitem[\protect\citeauthoryear{Gouldsbrough et~al.}{2022}]{GOULD2022}
\begin{botherref}
\oauthor{\bsnm{Gouldsbrough}, \binits{L.}},
\oauthor{\bsnm{Hossaini}, \binits{R.}},
\oauthor{\bsnm{Eastoe}, \binits{E.}},
\oauthor{\bsnm{Young}, \binits{P.J.}}:
{A temperature dependent extreme value analysis of UK surface ozone, 1980-2019}.
Atmospheric Environment
\textbf{273}
(2022)
\end{botherref}
\endbibitem

\bibitem[\protect\citeauthoryear{Hastie et~al.}{2009}]{hastibfri2009}
\begin{bbook}
\bauthor{\bsnm{Hastie}, \binits{T.}},
\bauthor{\bsnm{Tibshirani}, \binits{R.}},
\bauthor{\bsnm{Friedman}, \binits{J.}}:
\bbtitle{The Elements of Statistical Learning: Data Mining, Inference and Prediction}.
\bpublisher{Springer}, \blocation{New York}
(\byear{2009})
\end{bbook}
\endbibitem

\bibitem[\protect\citeauthoryear{Katz}{1999}]{katz99}
\begin{barticle}
\bauthor{\bsnm{Katz}, \binits{R.W.}}:
\batitle{Extreme value theory for precipitation: sensitivity analysis for climate change}.
\bjtitle{Advances in Water Resources}
\bvolume{23}(\bissue{2}),
\bfpage{133}--\blpage{139}
(\byear{1999})
\end{barticle}
\endbibitem

\bibitem[\protect\citeauthoryear{Katz et~al.}{2002}]{KATZ2002}
\begin{barticle}
\bauthor{\bsnm{Katz}, \binits{R.W.}},
\bauthor{\bsnm{Parlange}, \binits{M.B.}},
\bauthor{\bsnm{Naveau}, \binits{P.}}:
\batitle{Statistics of extremes in hydrology}.
\bjtitle{Advances in Water Resources}
\bvolume{25}(\bissue{8}),
\bfpage{1287}--\blpage{1304}
(\byear{2002})
\end{barticle}
\endbibitem

\bibitem[\protect\citeauthoryear{Kaufman and Rousseeuw}{1990}]{kaufman1990}
\begin{bbook}
\bauthor{\bsnm{Kaufman}, \binits{L.}},
\bauthor{\bsnm{Rousseeuw}, \binits{P.J.}}:
\bbtitle{Finding Groups in Data}.
\bsertitle{Wiley Series in Probability and Statistics}.
\bpublisher{John Wiley \& Sons, Inc.},
\blocation{Hoboken, NJ, USA}
(\byear{1990})
\end{bbook}
\endbibitem

\bibitem[\protect\citeauthoryear{Kiriliouk and Zhou}{2022}]{kiriliouk2022}
\begin{botherref}
\oauthor{\bsnm{Kiriliouk}, \binits{A.}},
\oauthor{\bsnm{Zhou}, \binits{C.}}:
Estimating probabilities of multivariate failure sets based on pairwise tail dependence coefficients
(2022).
Preprint at \url{https://arxiv.org/abs/quant-ph/0208066v1}
\end{botherref}
\endbibitem

\bibitem[\protect\citeauthoryear{{Klüppelberg} and Krali}{2021}]{klüppelberg2021}
\begin{barticle}
\bauthor{\bsnm{{Klüppelberg}}, \binits{C.}},
\bauthor{\bsnm{Krali}, \binits{M.}}:
\batitle{Estimating an extreme bayesian network via scalings}.
\bjtitle{Journal of Multivariate Analysis}
\bvolume{181},
\bfpage{104672}
(\byear{2021})
\end{barticle}
\endbibitem

\bibitem[\protect\citeauthoryear{Kunz et~al.}{2010}]{Kunz2010}
\begin{barticle}
\bauthor{\bsnm{Kunz}, \binits{M.}},
\bauthor{\bsnm{Mohr}, \binits{S.}},
\bauthor{\bsnm{Rauthe}, \binits{M.}},
\bauthor{\bsnm{Lux}, \binits{R.}},
\bauthor{\bsnm{Kottmeier}, \binits{C.}}:
\batitle{Assessment of extreme wind speeds from regional climate models – part 1: Estimation of return values and their evaluation}.
\bjtitle{Nat. Hazards Earth Syst. Sci.}
\bvolume{10},
\bfpage{907}--\blpage{922}
(\byear{2010})
\end{barticle}
\endbibitem

\bibitem[\protect\citeauthoryear{McLachlan and Peel}{2000}]{mclachlan2000}
\begin{bbook}
\bauthor{\bsnm{McLachlan}, \binits{G.}},
\bauthor{\bsnm{Peel}, \binits{D.}}:
\bbtitle{Finite Mixture Models}.
\bpublisher{Wiley Series in Probability and Statistics, John Wiley \& Sons, Inc.},
\blocation{New York}
(\byear{2000})
\end{bbook}
\endbibitem

\bibitem[\protect\citeauthoryear{Meyer and Wintenberger}{2023}]{meyer2023}
\begin{botherref}
\oauthor{\bsnm{Meyer}, \binits{N.}},
\oauthor{\bsnm{Wintenberger}, \binits{O.}}:
Multivariate sparse clustering for extremes.
Journal of the American Statistical Association,
1--12
(2023)
\end{botherref}
\endbibitem

\bibitem[\protect\citeauthoryear{Pickands}{1975}]{pickands75}
\begin{barticle}
\bauthor{\bsnm{Pickands}, \binits{J.}}:
\batitle{Statistical inference using extreme order statistics}.
\bjtitle{The Annals of Statistics}
\bvolume{3}(\bissue{1}),
\bfpage{119}--\blpage{131}
(\byear{1975})
\end{barticle}
\endbibitem

\bibitem[\protect\citeauthoryear{Resnick}{2004}]{resnick2004}
\begin{barticle}
\bauthor{\bsnm{Resnick}, \binits{S.}}:
\batitle{The extremal dependence measure and asymptotic independence}.
\bjtitle{Stochastic Models}
\bvolume{20}(\bissue{2}),
\bfpage{205}--\blpage{227}
(\byear{2004})
\end{barticle}
\endbibitem

\bibitem[\protect\citeauthoryear{Resnick}{2007}]{resnick2007}
\begin{bbook}
\bauthor{\bsnm{Resnick}, \binits{S.}}:
\bbtitle{Heavy-tail Phenomena: Probabilistic and Statistical Modeling}.
\bsertitle{Springer series in operations research and financial engineering}.
\bpublisher{Springer},
\blocation{New York, N.Y}
(\byear{2007})
\end{bbook}
\endbibitem

\bibitem[\protect\citeauthoryear{Rohrbeck et~al.}{2023}]{Rohr23}
\begin{botherref}
\oauthor{\bsnm{Rohrbeck}, \binits{C.}},
\oauthor{\bsnm{Simpson}, \binits{E.S.}},
\oauthor{\bsnm{Tawn}, \binits{J.A.}}:
Editorial: EVA 2023 data challenge
(2023)
\end{botherref}
\endbibitem

\bibitem[\protect\citeauthoryear{Towler et~al.}{2010}]{Towler10}
\begin{botherref}
\oauthor{\bsnm{Towler}, \binits{E.L.}},
\oauthor{\bsnm{Rajagopalan}, \binits{B.}},
\oauthor{\bsnm{Gilleland}, \binits{E.}},
\oauthor{\bsnm{Summers}, \binits{R.S.}},
\oauthor{\bsnm{Yates}, \binits{D.}},
\oauthor{\bsnm{Katz}, \binits{R.W.}}:
Modeling hydrologic and water quality extremes in a changing climate: A statistical approach based on extreme value theory.
Water Resources Research
\textbf{46}
(2010)
\end{botherref}
\endbibitem

\end{thebibliography}
\end{document}